\documentclass[referee]{jfm}

\usepackage{graphicx}
\usepackage{subfigure}
\usepackage{amssymb}
\usepackage{amsmath}
\usepackage{mathrsfs}
\usepackage{placeins}
\usepackage{natbib}
\usepackage{caption3}
\bibpunct{(}{)}{,}{s}{,}{,}

\usepackage[bookmarksnumbered=true,breaklinks=true,colorlinks,citecolor=blue,linkcolor=blue]{hyperref}

\newcommand{\Karman}{K\'arm\'an}

\makeatletter

\newcommand{\numtoRoman}[1]{\expandafter\@slowromancap\romannumeral #1@}

\renewcommand{\figurename}{Figure}
\usepackage[usenames]{color}

\makeatother

\begin{document}
\title[Exact coherent states in Poiseuille flow]{Exact coherent states and connections to turbulent dynamics in minimal channel flow}
\author[Jae Sung Park and Michael D. Graham]{Jae Sung Park and Michael D. Graham\thanks{Email address for correspondence: mdgraham@wisc.edu}}
\affiliation{%
Department of Chemical and Biological Engineering\\
University of Wisconsin-Madison, Madison, WI 53706-1691
\label{UWMad}
}%

\date{\today}

\maketitle

\begin{abstract}
Several new families of nonlinear three-dimensional travelling wave solutions to the Navier-Stokes equation, also known as exact coherent states, are computed for Newtonian plane Poiseuille flow. The symmetries and streak/vortex structures are reported and their possible connections to critical layer dynamics examined. While some of the solutions clearly display fluctuations that are localized around the critical layer (the surface on which the streamwise velocity matches the wave speed of the solution), for others this connection is not as clear. Dynamical trajectories along unstable directions of the solutions are computed. Over certain ranges of Reynolds number, two solution families are shown to lie on the basin boundary between laminar and turbulent flow. Direct comparison of nonlinear travelling wave solutions to turbulent flow in the same channel is presented. The state-space dynamics of the turbulent flow are organized around one of the newly-identified travelling wave families, and in particular the lower branch solutions of this family are closely approached during transient excursions away from the dominant behaviour. These observations provide a firm dynamical-systems foundation for prior observations that minimal channel turbulence displays time intervals of ``active'' turbulence punctuated by brief periods of ``hibernation'' (see e.g.~Xi, L. and Graham, M.~D., \emph{Phys.~Rev.~Lett.}, \textbf{104}, 218301 (2010)).  The hibernating intervals are approaches to lower branch nonlinear travelling waves.  Representing these solutions on a Prandtl-von K\'arm\'an plot illustrates how their bulk flow properties are related to those of Newtonian turbulence as well as the universal asymptotic state called maximum drag reduction (MDR) found in viscoelastic turbulent flow. In particular, the lower and upper branch solutions of the  family around which the minimal channel dynamics are organized appear to approach the MDR asymptote and the classical Newtonian result, respectively, both in terms of bulk velocity and mean velocity profile. 

\end{abstract}

\section{Introduction}
The understanding of the nature of near-wall turbulence has been greatly advanced by recent applications of dynamical systems theory to turbulent flow  \citep{Kawahara2012arfm}. In particular, over the past two decades, the discovery of three-dimensional fully nonlinear travelling wave (TW) solutions to the Navier-Stokes equations has enabled $\textit{a priori}$ study of self-sustained near-wall coherent structures that resemble in many ways that transient structures observed in fully turbulent flows  \citep{Hof04}. These solutions, also denoted as exact coherent states  \citep{Waleffe01} (ECS), are steady states in a reference frame translating at a constant streamwise speed. They have been found numerically in all canonical wall-bounded geometries for turbulent flows (plane Couette and Poiseuille, pipe and boundary layer)  \citep{Nagata90jfm, Clever97jfm, Nagata97pre, Waleffe98prl, Waleffe01, Waleffe03, Faisst03, Wedin04, Gibson08, Gibson09jfm, Schneider10prl, Duguet12prl, Blackburn13jfm}. Most solutions that have been found to date are spatially extended, but recent studies show the existence of spatially-localized travelling solutions that closely resemble turbulent puffs in the pipe flow geometry \citep{Avila13prl, Chantry14prl} or turbulent spots in the plane Couette and Poiseuille geometries \citep{Tillmark92jfm, Barkley05prl, Lemoult13jfm, Brand14jfm}. Other exact coherent states or other types of invariant solutions to the governing equation have been also found numerically. Periodic or relative periodic orbits were computed for plane Couette \citep{Kawahara01jfm, Viswanath07jfm, Cvitanovi10} and {Poiseuille flow \citep{Toh03jfm}}, pipe flow \citep{Duguet08jfm} and two-dimensional Kolmogorov flow \citep{Chandler13jfm}. Taken together, these solutions seem to form at least in part the dynamical skeleton underlying the chaotic dynamics of turbulent flow \citep{Gibson08, Kawahara2012arfm}. In the present work, we report and analyze several new families of such solutions in the plane Poiseuille geometry and further develop the understanding of the relationship between these solutions and the dynamics of turbulence. 

We focus here on plane Poiseuille (channel) flow of a Newtonian fluid with dynamic viscosity $\mu$, density $\rho$ and kinematic viscosity $\nu = \mu / \rho$ in a channel of half-height $h$. Characteristic inner scales are the friction velocity $u_{\tau}=(\tau_w/\rho)^{1/2}$ and the near-wall length scale or wall unit $\delta_{\nu} = \nu/u_{\tau}$, where $\tau_{w}$ is the time- and area-averaged wall shear stress. As usual, quantities nondimensionalized by these scales are denoted with a superscript "+". The friction Reynolds number is defined as $Re_{\tau}=u_{\tau}h/\nu=h/\delta_{\nu}$.

\begin{figure}
 \begin{center}\includegraphics[width=3.5in]{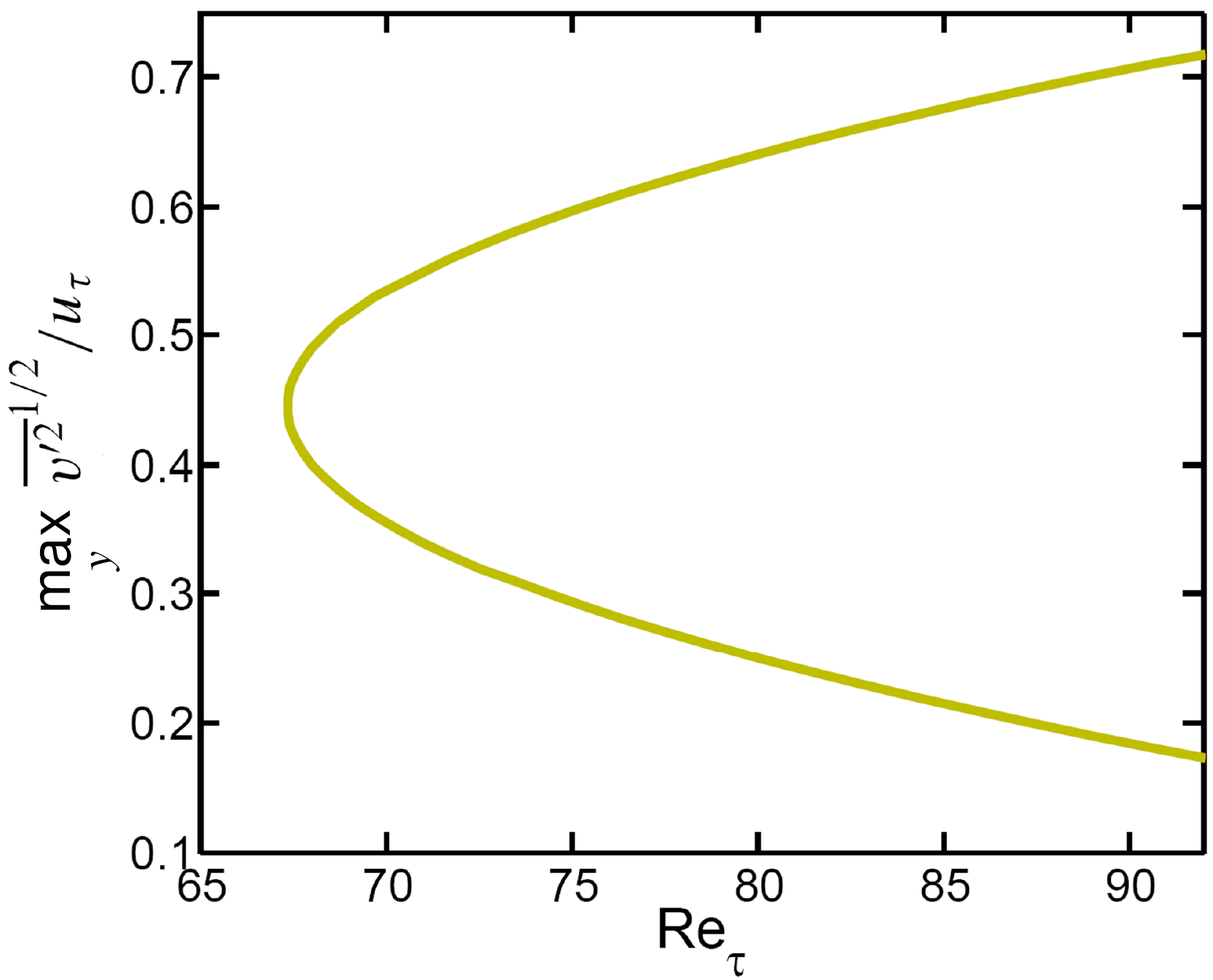}
  \caption{(Colour online) A bifurcation diagram for {one solution family of travelling waves found in the present study (labelled P4 below)}, where the maximum in the root mean square wall-normal fluctuations is shown against Reynolds number. \label{fig:intro}}\end{center}
\end{figure}

Exact coherent states {primarily} arise in pairs via a saddle-node bifurcation at a particular Reynolds number. At the bifurcation, the pair of solutions emerges at finite amplitude; we refer to each such pair as a solution family. 
Figure \ref{fig:intro} illustrates a bifurcation diagram of solution amplitude \emph{versus} Reynolds number for one such family (the ``P4'' family described below), using {the maximum over $y$ of} the root mean square wall-normal velocity fluctuations {$\overline{v'^2}^{1/2}$} normalized by the friction velocity {$u_{\tau}$} as a measure of solution amplitude. {An overbar indicates averaging over the streamwise and spanwise directions.} The so-called lower branch (LB) solution of each pair denotes a low-drag state owing to its lower maximum wall-normal velocity fluctuation compared to its corresponding upper branch (UB) solution. {(Additional solution branches can and do bifurcate off these primary states -- the ``P2'' solutions described below are one such example.)} In general, these solutions have a spatial structure in the form of low-speed streaks that are wavy in the streamwise direction, straddled by counter-rotating streamwise-aligned vortices: that is, they have the same basic qualitative structure as near-wall turbulence.

{The basic self-sustaining process underlying these structures has been qualitatively described by \cite{Waleffe97}. More recently it has been observed for Couette flow that at least one lower branch solution family has a structure that consists of streaks, rolls and a weak streamwise-varying wave that develops a critical layer -- i.e. its structure is localized around the surface where the streamwise velocity equals the wave speed of the ECS \citep{Wang07}. In the classical linear stability theory of parallel shear flows, a critical layer is a planar surface around which normal mode perturbations localize \citep{Drazin:1981wx}, while here the critical layer is a curved surface in 3D. \cite{Wang07} presented a scaling analysis suggesting that the wavy fluctuations should be localized in a region of thickness $O(Re^{-1/3})$ and showed that this scaling was followed by their numerical solutions. In fact, they found that the flow structures at $Re=50000$ and $Re=3000$ were virtually identical modulo a $Re^{1/3}$ rescaling of the direction normal to the critical layer surface. Hall and coworkers \citep{Hall10jfm,Blackburn13jfm} used a mixture of asymptotics and numerics to show, again for Couette flow, that the critical layer fluctuations couple back to the streamwise rolls to generate the nonlinear self-sustaining process that supports exact coherent states. In their formulation, this process is a version of wave-vortex interaction. They note that ``remarkable'' agreement is obtained between the high $Re$ asymptotics and the numerical results down to Reynolds numbers of order $10^{3}$ \citep{Hall10jfm}  -- we emphasize this point because this is the Reynolds number range of the present results. Other recent and interesting work on critical layers and ECS is found in \cite{Viswanath:2009da}, \cite{Deguchi:2013fr}, \cite{Gibson14jfm} and \cite{Deguchi:2014vx}.  Because of the clear importance of critical layer dynamics for at least some families of nonlinear travelling waves even at low $Re$, section 3.2 of the present work focuses on this topic.}

For plane Poiseuille flow, the first known families of travelling wave solutions were obtained by a homotopy continuation from ECS in plane Couette flow. Solutions were sought in a travelling reference frame using a Newton-Raphson method with the wave speed as an unknown \citep{Waleffe98prl, Waleffe01, Waleffe03, Nagata13, Gibson14jfm}. {Solutions have been also computed by a multiple shooting method \citep{Itano01}.}  It is worth noting that the lowest bifurcation point or onset Reynolds number for these solutions, $Re \approx 977$ or $Re_{\tau}\approx 44.3$, is in good quantitative agreement with the Reynolds number for transition to turbulence observed in an experiment in this geometry \citep{Carlson82jfm}. In fact, the solutions survive slightly below the critical Reynolds number for turbulence onset. Furthermore, the optimal spanwise wavelength $105.5\delta_{\nu} $ at the onset of travelling wave solutions is remarkably matched well with that of experimentally and numerically observed near-wall streak spacing of $80 - 120\delta_{\nu}$ \citep{Smith83jfm}. In the present work, several more families of travelling waves are found, which seem to have a closer connection to turbulent dynamics than the ones found earlier.

An important issue regarding exact coherent states is their connection to the laminar-turbulent boundary  -- the boundary in state space between the basins of attraction of the laminar and turbulent states. Some of the lower branch ECS found in turbulent shear flows are embedded on this boundary \citep{Itano01, Skufca06prl, Wang07, Schneider07prl, Kerswell07jfm, Duguet08jfm, Viswanath09jfm}. {In particular, such solutions that have only one unstable direction are called edge states \citep{Skufca06prl}.}  {Because such solutions are somehow the weakest, most marginal form of self-sustaining turbulence, the structure of the basin boundary and dynamical trajectories that lie on it are likely to play an important role in understanding the dynamics of transition to turbulence or onset of turbulence in wall-bounded shear flows. Recently, an experimental observation has been reported for the existence of edge states in pipe flow} \citep{Lozar12prl}.


Returning to the dynamical-systems point of view, exact coherent states are periodic (or more complicated but still invariant) orbits in state space, while the time evolution of a turbulent flow is a dynamical trajectory wandering around them. An important question is how closely the turbulent trajectories approach these invariant states. For plane Couette flow, \cite{Gibson08} visualized a clear illustration of this dynamical-systems viewpoint of turbulence by projecting the trajectory onto a set of orthonormal basis-states constructed with earlier ECS. \cite{Kerswell07jfm} also showed a clear visual illustration for a pipe flow. In addition to the state space visualization, they proposed quantitative measurements of the distance between a given instant on the turbulent trajectory and ECS, and suggested that ECS are visited for approximately 10 \% of the time in turbulent pipe flow. For a channel flow, the connections have yet to be fully made and will be investigated in the present study.

One important motivation for gaining a better understanding of turbulence is the possibility of reducing drag. In this context, the lower branch ECS are attractive due to their low-drag flow features and a natural question is whether it might be possible to able to somehow steer turbulent trajectories toward these states. One very successful approach to turbulent drag reduction is to add small amounts of rheologically active additives such as flexible long-chain polymers into a liquid \citep{White08arfm,Graham:2014uj}. The most dramatic effect of the polymer additives on turbulence occurs in the near-wall region, weakening the turbulent eddies in this region. The key feature of these polymer solutions in drag reduction is an existence of the so-called maximum drag reduction (MDR) phenomenon, at which very high levels of drag reduction are achieved by polymer additives, first identified by Virk \citep{Virk75}. The most intriguing observation for MDR is its universal mean velocity profile, the experimentally observed upper limit on the amount of drag reduction that can be achieved with polymer additives, also known as the Virk asymptote \citep{Virk75}. This asymptotic limit is insensitive to changes in the polymer solution such as concentration, molecular weight or polymer type. Thus, for a given situation, the maximum amount of drag reduction achievable with polymer additives is invariant. 

Li \emph{et al.}  \citep{Li06jfm, Li07pof} have investigated the effects of polymer additives on the channel flow ECS discovered by \cite{Waleffe01, Waleffe03}. For this solution family, as the level of viscoelasticity is increased, the Reynolds number for the solutions to come into existence increases. The primary effect of viscoelasticity on these ECS is the weakening of the streamwise vortices. Other effects are also seen in changes in the budgets of turbulent kinetic energy, Reynolds stress and the mean shear stress. All these effects show, at least at low levels of drag reduction, that the basic mechanism of drag reduction by polymers can be clearly elucidated by examining the impact of polymers on travelling wave solutions.  Nevertheless, these studies were limited to a single solution {family} and to relatively low Reynolds numbers and levels of viscoelasticity.

Another set of recent studies, while not directly focused on ECS, sheds some light on the state-space dynamics of Newtonian and viscoelastic channel flow. \cite{Xi:2010fv, Xi:2012io} performed DNS studies of minimal channel flow at low Reynolds numbers, finding that in Newtonian flow and for low to intermediate values of Weissenberg number, the flow cycles stochastically between ``active'' intervals, with strong streamwise vortices and three-dimensionality, velocity and a velocity profile near the von-\Karman{} profile, and ``hibernating intervals'', with very weak turbulence and a mean velocity profile approaching the Virk MDR profile. In viscoelastic flow, the polymers stretch during the active intervals, working against the streamwise vortices and shortening the duration of these intervals, while during the hibernating intervals the flow kinematics are very gentle and the polymers relax, only to begin stretching again at the beginning of the next active interval. Thus as the degree of viscoelasticity (Weissenberg number) increases, the overall dynamics look increasingly ``Virk-like'' as the active intervals contribute a decreasing time duration to the overall statistics.  A simple theory is developed, based on exponential stretching of polymers during active intervals and the idea that these intervals cannot persist once the polymer stress reaches a threshold value. This theory predicts the Weissenberg-number dependence of the duration of the active intervals in good agreement with the simulations. At high Weissenberg number, the hibernation intervals themselves are substantially altered and stabilized by viscoelasticity \citep{Wang:2014kn}, through mechanisms that are not yet understood. In addition, at low Reynolds number in the minimal channel flow geometry, hibernating turbulence is closely related to an edge state in which the Virk profile also arises, not just as a transient, but in the time-averaged velocity \citep{Xi12a}, even in Newtonian flow. A comprehensive overview of these studies is provided in \cite{Graham:2014uj}. The connection between active and hibernating turbulence and upper and lower branch Newtonian ECS will be solidified in the present work.

There are other indications as well that the Virk asymptote is not just universal for drag reduction by polymers but also arises in Newtonian turbulence. \cite{Dubief:2011uo} observed in a simulation of Newtonian boundary layer flow that at a spatial position just upstream of where vortices and turbulence spots form, the mean velocity profile looks strikingly similar to the Virk MDR profile.  Furthermore, the Virk MDR profile is also observed in a smoothed version of Newtonian plane Poiseuille flow in which spanwise length scales of the flow field below a specified size are suppressed \citep{Kerswell2003pof}. Finally, experimental observations of a Newtonian turbulent boundary layer flow subjected to spanwise wall oscillations display a mean velocity profile that, for $y^{+}\lesssim 30$, closely resembles the Virk MDR profile \citep{Bandyopadhyay:2006ji}. 

In this paper, we present five new families of nonlinear travelling wave solutions in Newtonian plane Poiseuille flow, and examine their spatiotemporal structure and connections to the dynamics of turbulent flow in the same geometry. In particular, we find a family whose upper and lower branch solutions have mean velocity profiles that resemble Newtonian turbulent (von K\'arm\'an) and MDR (Virk) profiles and we show the relationship between those solutions and trajectories of turbulent flows.  The problem formulation and solution methodologies are presented in Section \ref{sec:formulation}. Section \ref{sec:overview} presents an overview of mean flow properties and spatial structures of the solutions, while section \ref{sec:mode} illustrates the relation between the new travelling waves and critical layer dynamics. Sections \ref{sec:edge} and \ref{sec:connect} describe the connections between the travelling waves, the laminar-turbulent boundary and turbulent dynamics. Section \ref{sec:conclusion} presents conclusions.

\section{Formulation and solution approach\label{sec:formulation}}
We consider an incompressible Newtonian fluid in the plane Poiseuille geometry, driven with a constant mass flux $Q$. The characteristic length and velocity scales are the half-channel height $h$ and the laminar centerline velocity {$U_c = (3/4)Q/h$} for the same mass flux, respectively. With these characteristic scales, the Navier-Stokes equations in nondimensional form are
\begin{align}
&\nabla \cdot \boldsymbol{u}=0, \\
&\frac{\partial \boldsymbol{u}}{\partial t}+\boldsymbol{u}\cdot\nabla \boldsymbol{u}=-\nabla p+\frac{1}{Re_c}\nabla^2\boldsymbol{u}.
\end{align}
Here, we define the laminar equivalent Reynolds number for the given mass flux as {$Re_c = U_ch/\nu$}. {Note that $Re_b = U_b h/\nu = \frac{2}{3}Re_c$, where $U_b$ is the bulk velocity.} The $x, y$ and $z$ coordinates are aligned with the streamwise, wall-normal, and spanwise directions, respectively. Periodic boundary conditions are imposed in the $x$ and $z$ directions with fundamental periods $L_x$ and $L_z$, and no-slip conditions are imposed at the walls $y=\pm 1$. The computational domain is thus $[0, L_x]\times[-1,1]\times[0, L_z]$ or {simply $[L_x, 2, L_z]$}. The velocities are $u$, $v$, and $w$ in the $x$, $y$, and $z$ directions, and the velocity at point $(x,y,z)$ and time $t$ is expressed as $\boldsymbol{u}=[u,v,w](x,y,z,t)$.

Computation of nonlinear travelling waves is performed using the open source code \textit{ChannelFlow} written by \cite{Channelflow}, with use of a Newton-Krylov-hookstep algorithm \citep{Viswanath07jfm}. A numerical grid system is generated on $N_x \times N_y \times N_z$ (in $x$, $y$, and $z$) meshes, {where a Fourier-Chebyshev-Fourier spectral spatial discretization is applied to all variables.}  A typical resolution used is $(N_x, N_y, N_z) = (48, 81, 48)$. A travelling wave solution has the following form:
\begin{equation}
\boldsymbol{u}(x,y,z,t)=\boldsymbol{u}(x-c_x t,y,z), \label{eq:form}
\end{equation}
where $c_x$ is a constant wave speed in the streamwise direction. \textit{ChannelFlow} seeks solutions of a more general case:
\begin{equation}
\sigma \boldsymbol{f}^{t_1}(\boldsymbol{u})-\boldsymbol{u}=\boldsymbol{0}. \label{eq:basic}
\end{equation}
Here $\boldsymbol{f}^{t_1}$ is the time-$t_1$ forward time integration of the Navier-Stokes equations computed by a direct numerical simulation (DNS), i.e. $\boldsymbol{f}^{t_1}(\boldsymbol{u}(t))= \boldsymbol{u}(t+t_1)$  and  $\sigma$ is a symmetry operator to the flow field such that
\begin{equation}
\sigma[u,v,w](x,y,z)=[s_x u, s_y v, s_z w](s_x x + a_x L_x, s_y y, s_z z + a_z L_z). \label{eq:sym}
\end{equation}
Here we are following notations for flow symmetries introduced by \cite{Gibson08}. The symmetry operator $\sigma$ consists of two sets of parameters: $s_x, s_y, s_z$ for rotation-reflection symmetries (values are either 1 or -1) and $a_x, a_z$ for streamwise and spanwise translations (values are real). The symmetry operator $\sigma$ in (\ref{eq:basic}) describes the translation symmetry of the travelling wave solution after time $t_1$. To compute travelling wave solutions propagating in the streamwise direction, the only unknown symmetry parameter is the streamwise shift $a_x(=c_{x} t_{1}/L_{x})$, because, the spanwise shift $a_z$ is set to zero, the other symmetry parameters are inherent to the solution and the time shift $t_1$ is chosen \emph{a priori}. The parameter $a_{x}$ is determined as part of the solution process.

More generally, the symmetries of fluid states can be expressed with the symmetry operator (\ref{eq:sym}). That is, $\boldsymbol{u} = \sigma \boldsymbol{u}$ for certain values of symmetry parameters. The symmetry operator $\sigma$ is then expressed in different characters to describe different symmetries of fluid states: $\tau$ for the spatial phase shifts, $\sigma$ for the reflections and $s$ for the shift-reflection or shift-rotation. The four flow symmetries that arise in the present study are:
\begin{align}
&\sigma_y[u,v,w](x,y,z)=[u,-v,w](x, -y, z), \\
&\sigma_z[u,v,w](x,y,z)=[u,v,-w](x, y, -z), \\
&\tau_{xz}[u,v,w](x,y,z)=[u,v,w](x+\frac{L_x}{2}, y, z+\frac{L_z}{2}) \\
&s_{1}[u,v,w](x,y,z)=[u,v,-w](x+\frac{L_x}{2}, y, -z).
\end{align}
The $\sigma_y$ and $\sigma_z$ symmetries correspond to reflections with respect to the midplanes in the $y$ and $z$ directions, respectively. The $\tau_{xz}$ and $s_1$ symmetries denote half-domain translations in the $x$ and $z$ directions and a shift-reflection symmetry, respectively. In particular, the $s_1$ symmetry is related to the sinusoidal instability of streaks \citep{Waleffe97}, which is also called the fundamental sinuous mode.


Finding solutions to equation~(\ref{eq:basic}) requires good initial guesses. We generate these using instantaneous velocity fields from DNS of turbulent trajectories that have been symmetrized with respect to the midplane of the domain, $y=0$ (i.e.~all initial guesses satisfy $\boldsymbol{u}=\sigma_{y}\boldsymbol{u}$). In particular, since hibernating turbulence has been hypothesized to be closely related to travelling wave solutions \citep{Xi12a}, we chose initial guesses from instants with a lower wall shear stress than the mean value. {The time shift $t_1$ is arbitrary. A relatively large value provides substantial improvement to the rate of convergence of the Krylov subspace methods that are used in our computation, but larger values of $t_1$ require longer to compute $\boldsymbol{f}^{t_1}$. We chose $t_1=20$ in that it seems to balance these two aspects well. The initial guess for the streamwise shift $a_x$ is determined by approximating the wave speed as the bulk velocity of the symmetrized initial velocity field.} {Appropriate symmetries are enforced during the time-$t_1$ time integration and the search procedure.} To solve equation~(\ref{eq:basic}) a Krylov subspace method is used to solve the linear systems arising at each Newton step. For better convergence, a trust-region limitation to the magnitude of the Newton steps or a hook step within a Krylov subspace is computed for the optimal Newton step. The Newton iteration is repeated until an accuracy of $O(10^{-15})$ is reached, {where the accuracy is the residual of $\parallel \sigma\boldsymbol{f}^{t_1}(\boldsymbol{u}) - \boldsymbol{u} \parallel$, using the $L_2$ norm
\begin{equation}
\parallel \boldsymbol{g} \parallel = \left[ \frac{1}{2L_xL_z} \int_0^{L_z}\int_{-1}^1\int_0^{L_x} \boldsymbol{g\cdot g}~\textrm{d}x \textrm{d}y \textrm{d}z \right]^{1/2}.
\end{equation}
Note that the \textit{ChannelFlow} code calculates this residual.} 
A detailed description for the solution algorithm and DNS can be found in \cite{Gibson08, Gibson09jfm} and \cite{Viswanath07jfm,Viswanath:2009da}.

\begin{table}
\begin{center}
  \begin{tabular}{c c c c c c c c}
    \hline
    TW & $Re_c$ & & $Re_{\tau}$ & $L_x $ & $L_z $ & $c_x$ & Symmetries \\ \hline
    P1 & 1315.96 & &58.95 & $\pi$ & $\pi$   & 0.73  & $\sigma_y, \sigma_z, \tau_{xz}$ \\ \hline
    P2 & 1270.98 & &59.88 & $\pi$ & $\pi$   & 0.71  & $\sigma_y$ \\ \hline
    P3 & 682.57 & & 43.20 & $2\pi$ & $\pi$   & 0.62 & $\sigma_y, \tau_{xz}, s_1$ \\ \hline
    P4 & 1400 & & 67.32 & $\pi$ & $\pi/2$ &  0.75 & $\sigma_y, \sigma_z$ \\ \hline
    P4SB & 3070 & & 88.70 & $2\pi$ & $\pi$ &  0.78 & $\sigma_y, \sigma_z, \tau_{xz}$ \\ \hline
    P5 & 2744.95 & & 99.02 & $\pi$ & $\pi/2$ & 0.78  & $\sigma_y, \sigma_z$ \\ \hline
  \end{tabular}
  \caption{Scales and symmetries of travelling wave solutions at their saddle-node bifurcation points. The wave speed is normalized by laminar centerline velocity. Note that the bifurcation points for P2 and P4 subharmonic branch (SB) correspond to its minimum due to the discovery of only one branch. \label{table1}}
\end{center}
\end{table}

\begin{figure}
 \begin{center}\includegraphics[width=5.5in]{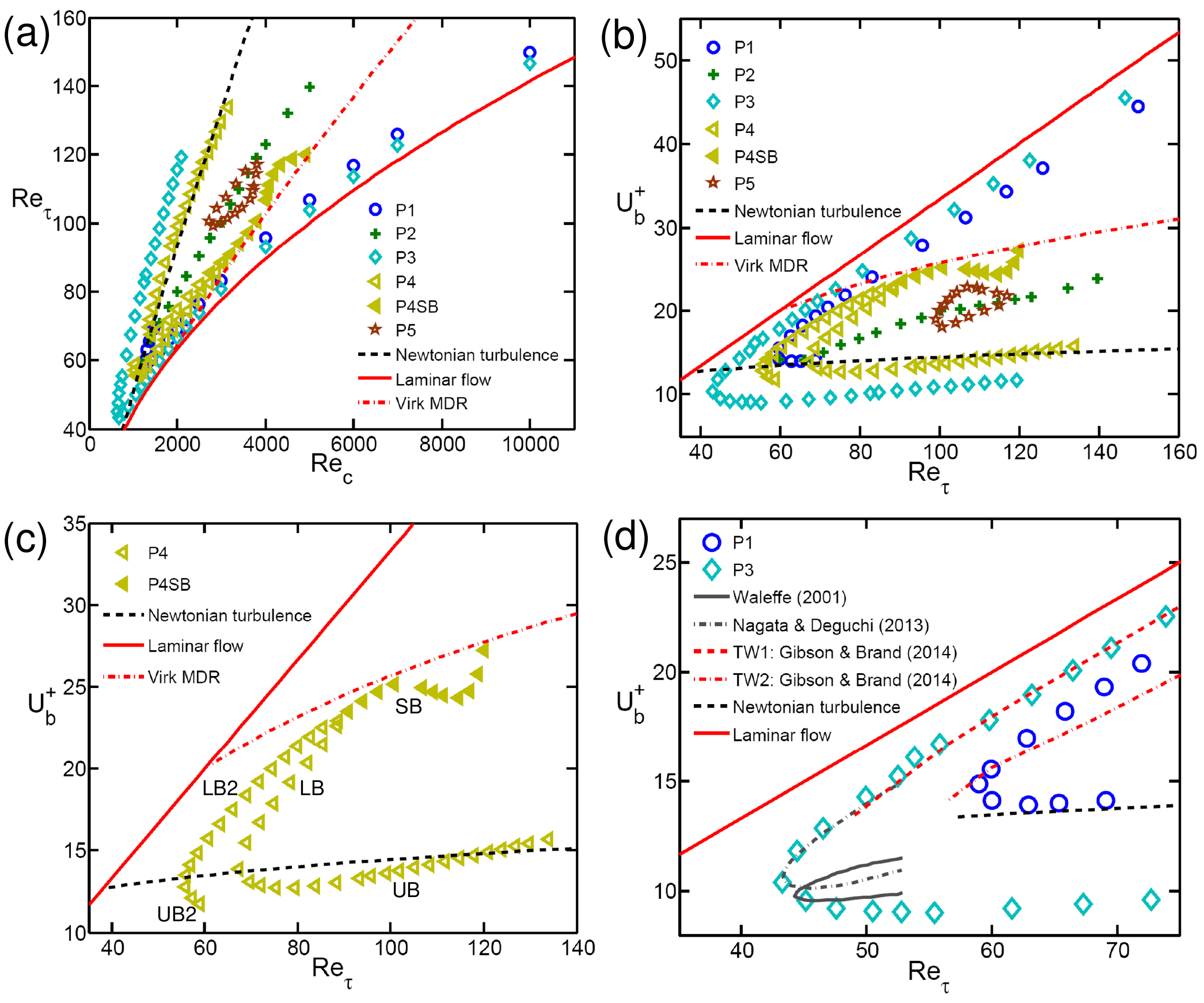}
  \caption{(Colour online) (a) A bifurcation diagram for five families of travelling wave solutions in terms of Reynolds number $Re_c = U_ch/\nu$ and friction Reynolds number $Re_{\tau} = u_{\tau} h/\nu$. The curves for Newtonian turbulence and laminar flows are shown. For P4, solid symbols correspond to solutions computed by a subharmonic bifurcation (SB). (b) A Prandtl-von K$\acute{\textrm{a}}$rm$\acute{\textrm{a}}$n plot for a bifurcation diagram. The average velocities as a function of friction Reynolds number are shown along with curves for Newtonian turbulence, laminar flow and the Virk maximum drag reduction. (c) Only P4 solution family is shown. (d) A comparison of P1 and P3 to earlier solutions of \cite{Waleffe01}, \cite{Nagata13} and \cite{Gibson14jfm} in the same geometry.\label{fig:Uave}}\end{center}
  \end{figure}
  
\section{Results and discussion\label{sec:results}}
\subsection{Travelling wave families: bifurcation diagram, mean profiles and overall structures \label{sec:overview}}
We computed five families of nonlinear travelling wave solutions in plane Poiseuille flow, which we have labelled P1 through P5, in domains of three different sizes: $[\pi, 2, \pi]$, $[2\pi, 2, \pi]$, and $[\pi, 2, \pi/2]$. {For the same flow geometry, the previous studies of \cite{Nagata13} and \cite{Gibson14jfm} used the domain of $[2\pi, 2, \pi]$ and \cite{Toh03jfm} used the domain of $[\pi, 2, 0.4\pi]$. The minimum spanwise domain size used here is about $95\delta_{\nu}$, corresponding to the length scales of the optimal spanwise wavelength for ECS \citep{Waleffe03} and of the near-wall streak spacing of about $100\delta_{\nu}$ \citep{Smith83jfm}. This minimum spanwise length scale is also within the range of the critical channel widths about 85 -- 110 wall units used for the minimal flow unit studies \citep{JIMENEZ:1991tl}.}  
Table \ref{table1} presents scales and symmetries of the solutions at their bifurcation points. Since only one solution branch is found for P2, the lowest $\textit{Re}$ solution is presented. Because we imposed the $\sigma_y$ symmetry on initial guesses, all solutions exhibit this $\sigma_y$ symmetry; P2 has only this symmetry. The half-period translations in the streamwise and spanwise directions, $\tau_{xz}$, are found for P1 and P3. The shift-reflect symmetry $s_1$ responsible for the fundamental sinuous mode is found for P3. 

The bifurcation diagram for these solution families is shown in figure~\ref{fig:Uave}(a). The solutions are plotted using friction Reynolds number vs. laminar equivalent Reynolds number. For each solution family, a solution with higher $Re_{\tau}$ is an upper branch solution corresponding to high drag, while its counterpart is a lower branch solution. For comparison, Newtonian turbulence and laminar flow are also drawn. Another representation of the bifurcation diagram, a Prandtl-von \Karman{} plot, is shown in figure~\ref{fig:Uave}(b). This form is often used to represent drag-reduction characteristics in wall-bounded turbulent flows. {The bulk velocities $U^+_{b}$} are plotted as a function of friction Reynolds number along with curves for Newtonian turbulence, laminar flow and Virk MDR. The curve for the Virk MDR is generated using its universal mean velocity profile \citep{Virk75}. We elaborate below on the solutions with respect to the Prandtl-von \Karman{} plot. In this representation, a ``lower branch'' solution is above the ``upper branch'', because the former has higher bulk velocity for the same wall shear stress than the latter. With the exception of the lower branch solutions of P1 and P3, the maximum $Re$ at which a solution is shown on the bifurcation diagram represents the highest $Re$ at which a converged solution could be found. Obtaining solutions at higher $Re$ will require further refinements in techniques for solving equation~(\ref{eq:basic}).

The lower branches of P1 and P3 become very close and parallel to the laminar solution as Reynolds number increases. Their closeness to the laminar state indicates that they are very low drag states. These lower branch solutions have very weak spatial variations and Reynolds number dependence and have been successfully continued up to $Re_{\tau} \approx 300$ (corresponding to $Re_c \approx 40000$). Regarding the upper branches of the solution families, in the range where we have computed it P1 has a similar level of drag (i.e. a similar bulk velocity for a given $Re_{\tau}$) as Newtonian turbulence. The P3 upper branch, however, shows higher drag than Newtonian turbulence, displaying the highest drag level among the solutions found in the present study. 
The P2 solution branch appears to bifurcate off P1, a result that is confirmed below when we see that P2 has a broken $\sigma_{z}$ symmetry.  P5 forms a closed loop (isola). 


Let us now focus on the P4 solution family, which shows very intriguing behaviour with regard to Newtonian and viscoelastic turbulence. Figure \ref{fig:Uave}(c) shows only the P4 solution family. Consider the upper branch (low velocity) solution at the upper range of convergence $Re_{\tau}\approx 130$. This solution branch has a mean velocity very close to that of Newtonian turbulence (the black dashed curve). As $Re$ decreases along this branch, the solution remains close to the Newtonian turbulence curve until undergoing a turning point at $Re_{\tau}=67.32$, beyond which the lower branch (LB) appears to approach the Virk MDR curve. This solution branch turns around again at $Re_{\tau}=88.7$, forming another lower branch solution (which we call LB2). As we decrease Reynolds number, the new branch reaches another bifurcation point at $Re_{\tau}=55.63$ and we denote the solution beyond this point as UB2. Thus, there are three bifurcation points for P4 solution family. Interestingly, the bulk velocities at the bifurcation points at $Re_{\tau}=55.63$ and $67.32$ are remarkably close to the Newtonian turbulence value, while the third bifurcation point at $Re_{\tau}=88.7$ is close to the Virk MDR value. Finally, above the turning point at $Re_{\tau}=88.7$, a subharmonic -- spatiotemporal period-doubling --  branch (SB) arises, which has doubled fundamental spatial periods in the $x$ and $z$ directions compared to the P4 solution family (i.e. $L_x$ becomes $2\pi$ and $L_z$ becomes $\pi$){, while the wave speed remains constant. Thus at any point in the domain the temporal period of the velocity, as measured in the laboratory frame, doubles.} The subharmonic solutions are indicated by solid symbols. This solution closely follows the MDR curve until $Re_{\tau} \approx 105$, deviating from and then returning to it as $Re$ increases further. 

Prior to proceeding to figure \ref{fig:Uave}(d), it is worth mentioning the linear stability of the solutions. The leading eigenvalues of the solutions are computed {in their symmetric subspace} with Arnoldi iteration \citep{Viswanath07jfm}. The P1, P3, and P5 lower branch solutions have a single, real unstable eigenvalue, while the P4 lower branch (P4-LB) solution has two real unstable eigenvalues. The P4-LB2 solution has three real unstable eigenvalues, and the P4 subharmonic branch (P4-SB) solution has three real and three complex conjugate unstable eigenvalues. Turning from the P4 lower branch to the P4 upper branch (or from P4-LB2 to P4-UB2), one real unstable eigenvalue goes complex immediately in a Takens-Bogdanov bifurcation  \citep{Guckenheimer:1983up}, at which an eigenvalue associated with a saddle-node bifurcation collides with another eigenvalue. So just beyond their respective turning points, P4-UB and UB2 have one real and one complex conjugate pair and two real and one complex conjugate pair of unstable eigenvalues, respectively. The P1, P3, and P5 solutions also experience the Takens-Bogdanov bifurcation  after crossing lower to upper branch. Interestingly, this behavior has also been observed near turning points of pipe flow travelling waves \citep{Mellibovsky2011jfm, Pringle:2009cja} and thus seems to be rather generic for travelling waves in shear flows. 

\begin{figure}
 \begin{center}\includegraphics[width=5.5in]{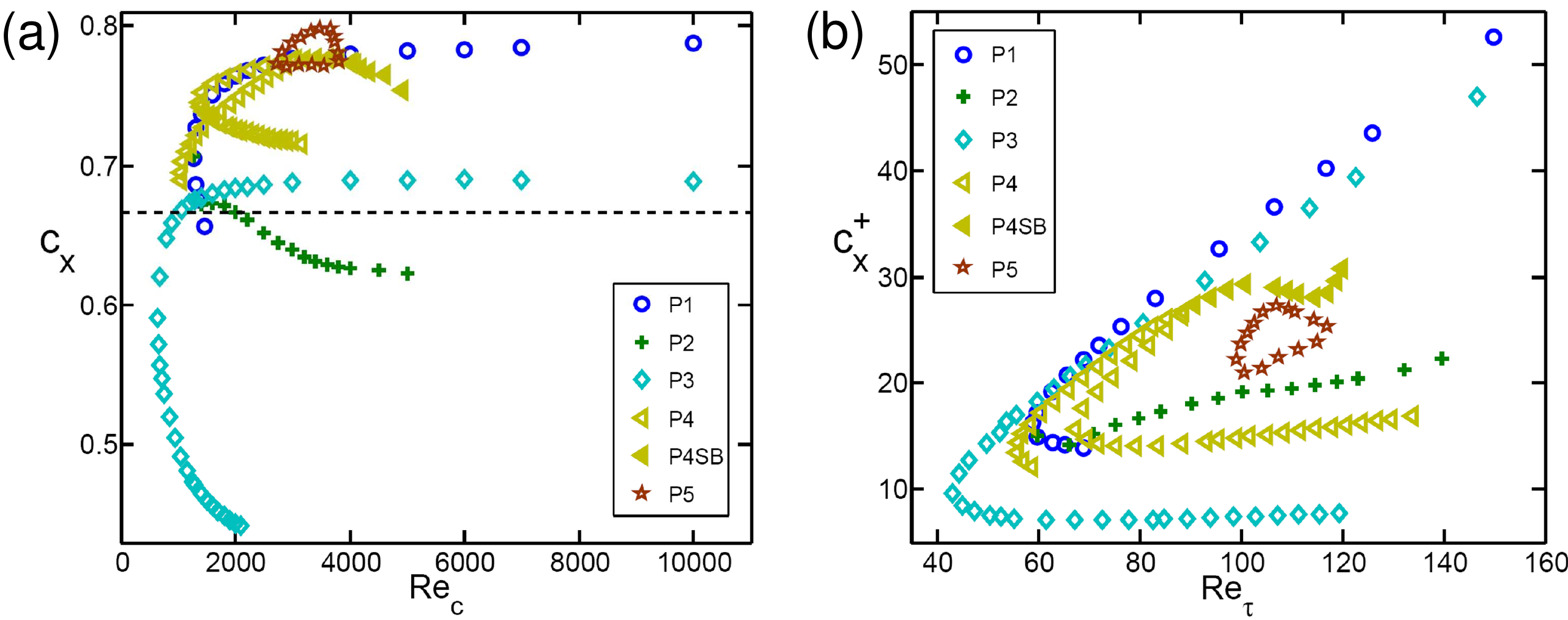}
  \caption{(Colour online) Wave speeds for the travelling waves as a function of Reynolds number in (a) outer units and (b) inner units. In (a), the wave speeds are normalized by the laminar centerline velocity and the dashed line represents the laminar bulk velocity.  \label{fig:cx}}\end{center}
\end{figure}

Figure \ref{fig:Uave}(d) compares P1 and P3 to earlier TW solutions discovered by \cite{Waleffe01}, \cite{Nagata13} and \cite{Gibson14jfm} in the same geometry. The curves for Waleffe and Nagata \& Deguchi solutions are generated from figure 5 of \cite{Nagata13}, where the optimal wavelengths were used. Gibson \& Brand's two solutions, named TW1 and TW2 in their paper, have the same wavelengths as P3. The solution of Nagata \& Deguchi, called MS-S in their paper, possesses $\sigma_y$, $\sigma_z$ and $s_1$ symmetries, and Waleffe solution lacks the $\sigma_z$ symmetry compared to MS-S solution. TW1 has the same symmetries as P1, while the $\sigma_y$ symmetry is lost in TW2. The bifurcation points of P3, Waleffe and MS-S solutions are very close each other. As shown, the P3 upper branch appears to be the highest drag state. Note that Waleffe, MS-S and TW1 solutions were obtained by continuation from plane Couette to Poiseuille conditions, but TW2 was obtained by a similar manner to an edge tracking method using a velocity field from DNS as an initial guess. To our knowledge, solutions similar to P4 and P5 have not previously been found.

\begin{figure}
 \begin{center}\includegraphics[width=5.5in]{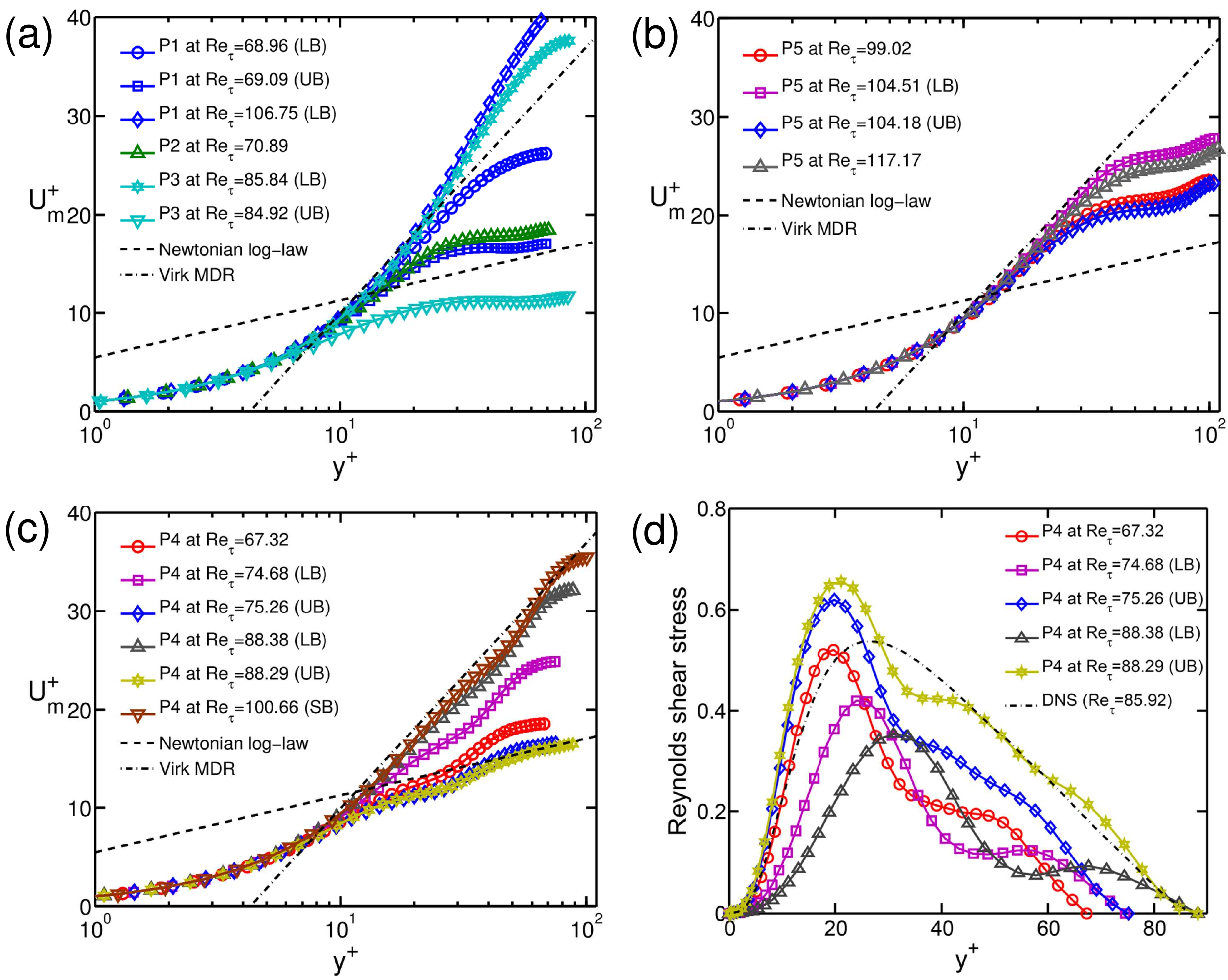}
  \caption{(Colour online) Mean velocity profiles for (a) P1, P2 and P3, (b) P5, and (c) P4, in comparison to the log-laws for Newtonian turbulence and Virk MDR. (d) Mean profiles of the Reynolds shear stress for P4 solution family. A dot-dashed line is a time-average profile for long DNS trajectories. \label{fig:um}}\end{center}
\end{figure}

Figures \ref{fig:cx}(a) and (b) show the wave speed $c_x$ of the solutions as a function of Reynolds number in outer and inner units, respectively. In general, the wave speed follows the same trend as the bulk velocity: in a given solution family, a lower branch solution propagates faster than an upper branch solution. In outer units in figure \ref{fig:cx}(a), the laminar bulk velocity is plotted in a dashed line. Most of solutions have a larger wave speed than the laminar bulk velocity, indicating that they propagate forward when viewed in a reference frame moving at the laminar bulk velocity. The wave speeds of the P1 and P3 lower branch solutions appears to become constant with increasing Reynolds number, while those of their upper branch solutions decrease drastically. When plotted in inner units, the wave speed behaviour shows almost the same shape as the bulk velocity plot in figure \ref{fig:Uave}.

We now turn our attention to the mean velocity profiles $U^+_{m}(y^+)$. Figure \ref{fig:um}$(a)$ shows these for P1, P2 and P3.  For comparison, we also plot the von \Karman{} log-law $U^+_{m}(y^+) = 2.5~\textrm{ln} y^+ + 5.5$ profile of Newtonian turbulence and the Virk log-law $U^+_{m}(y^+) = 11.7~\textrm{ln}y^+ - 17.0$ that approximates the mean velocity profile in the MDR regime. As expected from the average velocity results in figure \ref{fig:Uave}(b), the lower branch velocity profiles for P1 and P3 are well above the Virk MDR profile for $Re_{\tau} > 80$ and very close to the parabolic laminar profile. The highest-drag solution of the P3 upper branch solution shows a mean velocity profile well below the von \Karman{} log-law profile. The  velocity profile for P2 shows a similar character to the P1 and P3 upper branch profiles.

In figure \ref{fig:um}(b), the mean velocity profiles for P5 are shown at its minimum and maximum Reynolds numbers, and at $Re_{\tau} \approx 104$. In particular, the lower branch velocity profile at $Re_{\tau}=104.51$ very closely approaches the Virk MDR log-law in the range $15 < y^+ < 45$.


Figure \ref{fig:um}(c) shows mean velocity profiles for P4. Starting from its first bifurcation point  at $Re_{\tau}=67.32$, the upper and lower branch profiles seem to approach toward two distinct limits, the von \Karman{} and  Virk MDR profiles, respectively, as Reynolds number is increased. In particular, the lower branch profile at $Re_{\tau}=88.38$ very closely approaches the Virk MDR log-law profile over a relatively wide range, $15 < y^+ < 70$. This lower branch profile is very similar to the conditionally-sampled DNS velocity profile and experimentally observed profile for Newtonian hibernating turbulence \citep{Xi:2012io, Whalley14}. Meanwhile, {the upper branch profile lies very close to the experimentally-observed mean profile of Newtonian turbulent flows and approaches the von \Karman{} log law profile at large $y^+$}, as does active turbulence. The subharmonic branch profile is also presented for $Re_{\tau}=100.66$, showing a similar shape as the lower branch. Hence, the upper, lower and subharmonic branches of the P4 solution family may represent an envelope in state space encompassed by {the mean profiles of both the Virk MDR and the classical Newtonian turbulence}.

For the P4 solution family, we plot Reynolds shear stress profiles in figure \ref{fig:um}(d), in comparison to the time-average Newtonian profile for long DNS trajectories. As Reynold number is increased, the Reynolds shear stress for lower branch solutions decreases, while the upper branch profile increases. Compared with the time-average Newtonian profile at $Re_{\tau} \approx 86$, the reduction in the lower branch solution is substantial. Interestingly, the upper branch profile is slightly higher compared with magnitudes of the Newtonian profile in $y^+ < 30$, but it almost collapses onto the Newtonian profile for $y^+ > 45$.

\begin{figure}
 \begin{center}\includegraphics[width=5.5in]{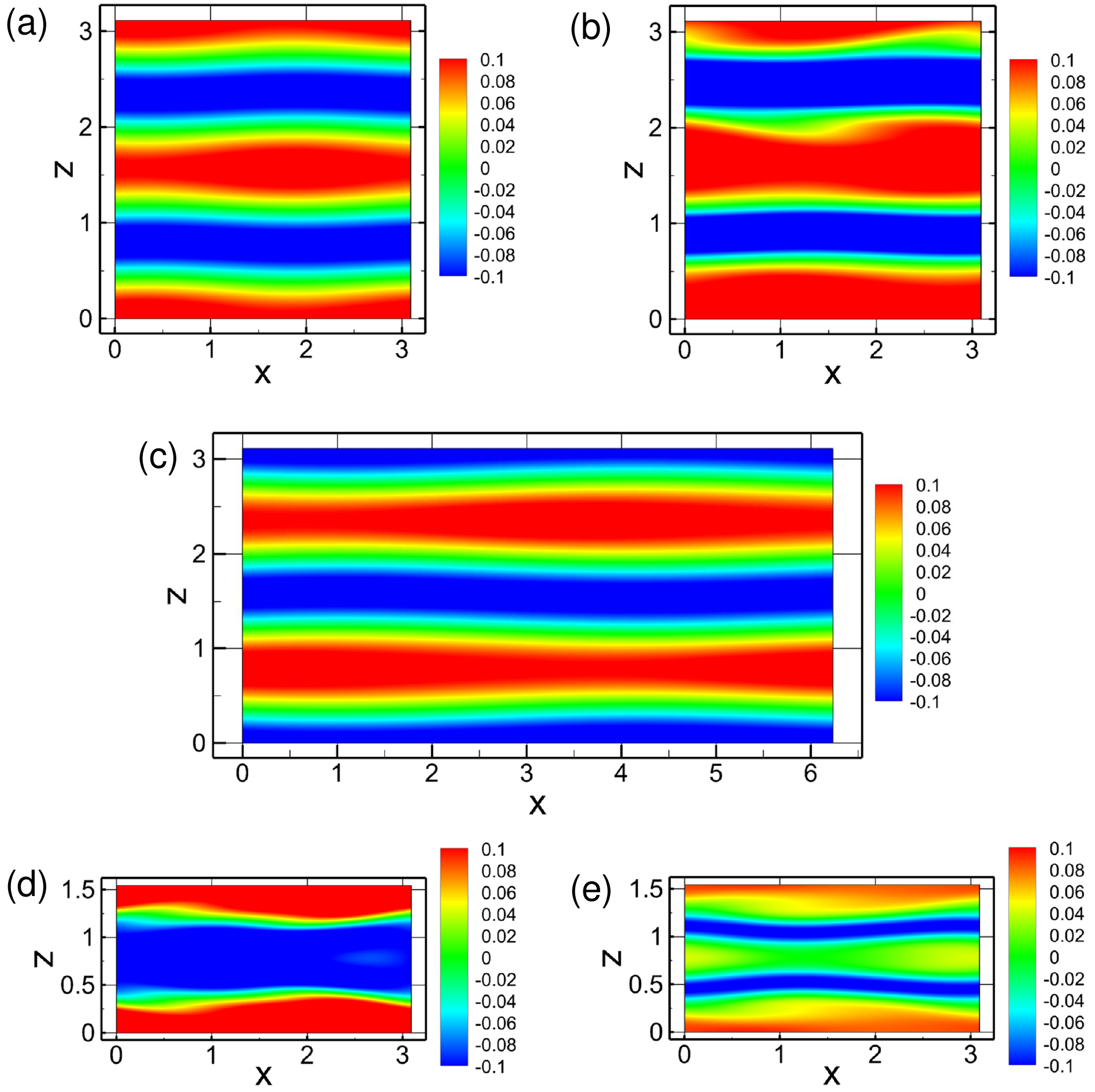}
  \caption{(Colour online) Contours of streamwise velocity fluctuation in the $x-z$ plane at $y = -0.5$ for (a) P1, (b) P2, (c) P3, (d) P4 at $Re_c = 1800$, and (e) P5 at $Re_c = 3600$. {Lower branch solutions are presented except for P2, for which there is only one branch}. Blue and red indicate low- and high-speed streaks, respectively. \label{fig:streak}}\end{center}
\end{figure}

\begin{figure}
 \begin{center}\includegraphics[width=4.0in]{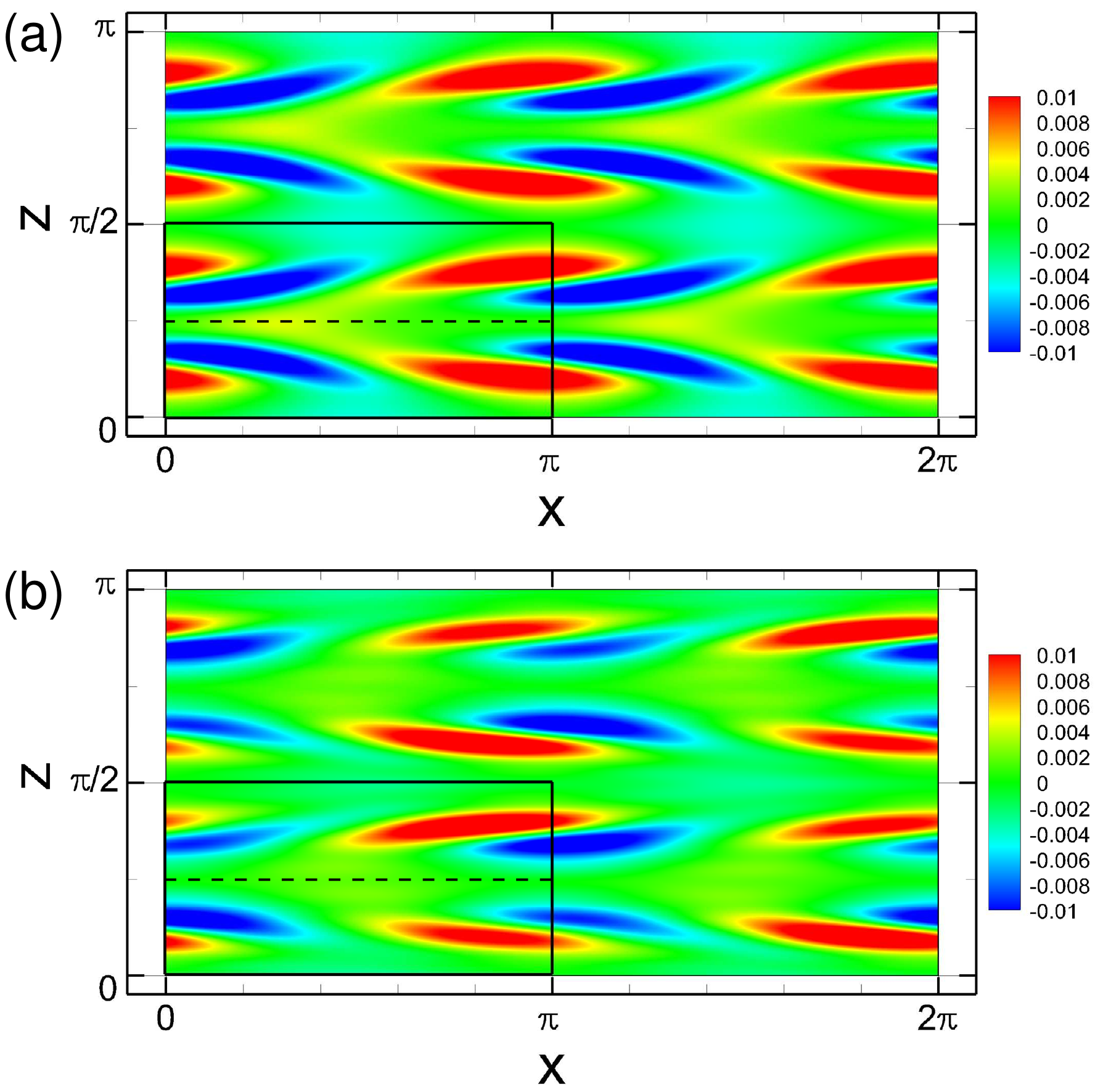}
  \caption{(Colour online) (a) P4 lower branch solution  at $Re_c = 2750$ (two periods in $x$ and $z$ are shown) and (b) subharmonic solution at $Re_c = 3800$. Shown are colour contours of the wall-normal velocity in the $x-z$ plane at $y = -0.5$. The solid-outlined box at lower left shows the size of the domain in which the P4 solutions are found. The subharmonic solution has a broken mirror-symmetry about the midplane of this box in the $z$ direction (dashed line), as well as a broken discrete translation symmetry in both the $x$ and $z$ directions -- the solution in $\pi<x<2\pi$ is shifted by $\pi$ in the $z$-direction from the solution in $0<x<\pi$. \label{fig:subharmonic}}\end{center}
\end{figure}

\begin{figure}
 \begin{center}\includegraphics[width=5.5in]{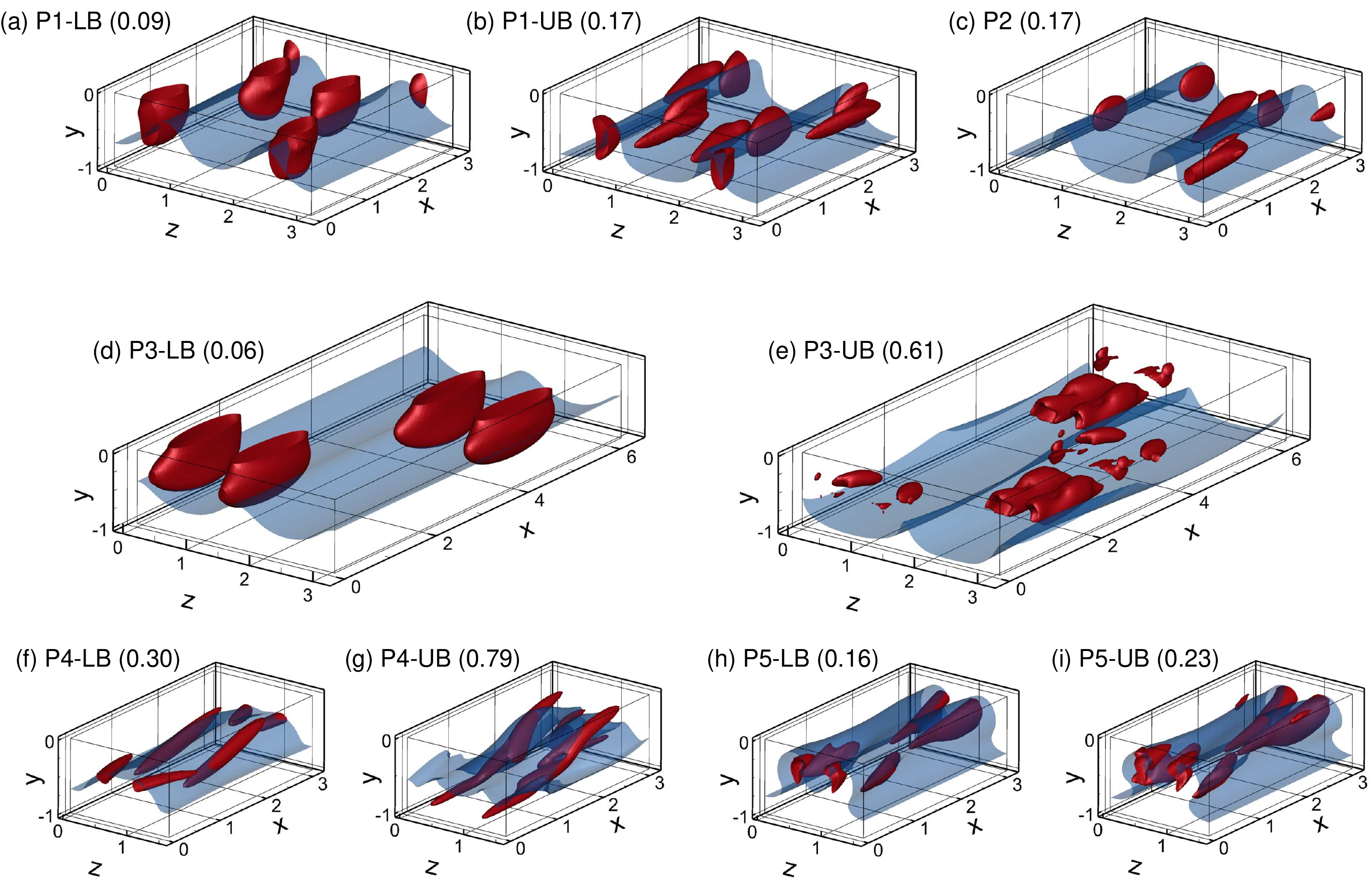}
  \caption{(Colour online) Vortical structures of the travelling wave solutions as  illustrated by the swirling strength $\lambda_{ci}$.  (a) lower and (b) upper branch solutions for P1, (c) P2 solution, (d) lower and (e) upper branch solutions for P3, (f) lower and (g) upper branch solutions for P4, (h) lower and (i) upper branch solutions for P5. P5 is shown at $Re_c = 3600$, all others at $Re_c = 1800$. The bottom-half channel is only shown due to the mirror-symmetry with respect to the channel center. The numbers in parenthesis are the maximum swirling strength. Dark red tubes are isosurfaces at $2/3$ of the maximum swirling strength, and transparent blue isosurfaces indicate critical layer surfaces, where the local streamwise velocity matches the wave speed. \label{fig:vortex}} \end{center}
\end{figure}

Now we examine the streak structures, as represented by the contours of streamwise velocity fluctuations in the $x-z$ plane at $y = -0.5$; these are shown in figure \ref{fig:streak}. Except for P2, lower branch solutions are presented. The low-speed and high-speed streaks are denoted by negative (blue) and positive (red) fluctuations, respectively. A subharmonic sinucose mode \citep{Waleffe97}, which is a combination of sinusoidal for the low-speed streak and a varicose mode for the high-speed streak,  is identified for P1, P3 and P5. A fundamental sinuous mode is observed for the P2 solution (figure \ref{fig:streak}(b)), while P4 exhibits a fundamental varicose mode (figure \ref{fig:streak}(d)). From a flow symmetry point of view, the $\sigma_z$ symmetry is clearly seen for P1, P4 and P5, whereas this symmetry is broken for P2 and P3. The $\tau_{xz}$ symmetry for P1 and P3 is also identified in figures \ref{fig:streak}(a) and (c).


To clearly illustrate the subharmonic bifurcation arising on the P4 lower branch around $Re_c=3070$, we plot in figures \ref{fig:subharmonic} (a) and (b) the wall-normal velocities on the $x$-$z$ plane of solutions at $Re_c = 2750$ and $3800$, respectively. The fundamental spatial periods of the P4 lower branch solution are $L_x=\pi$ and $L_z=\pi/2$; figure \ref{fig:subharmonic}(a) shows two periods of this solution in each direction -- the unit cell is the solid-outlined box at lower left. In the subharmonic solution, figure \ref{fig:subharmonic}(b), mirror-symmetry with respect to the $z$ direction midplane (dashed line) of the unit cell of P4 is broken, even though the $\sigma_z$ symmetry still holds for its own (larger) fundamental domain. In addition, the subharmonic solution has a broken discrete (half domain shift) translation symmetry in both $x$ and $z$ directions: the solid-outlined box is not same as the solution in $\pi < x < 2\pi$ and $0 < z < \pi/2$ or in $0 < x < \pi$ and $\pi/2 < z < \pi$.

The streak structure of a flow is closely related to the streamwise vortical structure. Figure \ref{fig:vortex} shows contours of swirling strength $\lambda_{ci}$, the imaginary part of the complex conjugate eigenvalues of the velocity gradient tensor \citep{Zhou1999jfm},  in the bottom half of the channel. The contours represent $2/3$ of the maximum swirling strength for each solution, which is presented in parentheses in the figure. In a given solution family, lower branch solutions have weaker vortex strength than upper branch solutions. We also depict the critical layer surface, where the local steamwise velocity equals the wave speed, $u(x,y,z,t)=c_x$ \citep{Maslowe86, Hall10jfm}. The P1 and P3 lower branch solutions, which have the subharmonic sinucose mode, show similar vortical structures. The vortex cores appear to expand between just above critical layer and the channel center, whereas the vortex cores of the P1 and P3 upper branch solutions are located very close to the critical layer. The fundamental sinuous mode of P2 displays staggered vortices, which are also located close to the critical layer. The fundamental varicose mode of the P4 lower and upper branch solutions displays different inclination angles of the vortex legs with respect to the wall: the vortex legs of the lower and upper branch solutions are inclined about 20 and 40 degrees to the wall, respectively. The vortex cores are also located around the critical layer. This vortical structure  -- which has the same symmetry as a hairpin but does not display a ``head'' -- is also observed for other travelling wave solutions in the same geometry \citep{Gibson14jfm} and plane Couette flow \citep{Itano09prl, Deguchi10pre}.

\subsection{Travelling wave structure and critical layers \label{sec:mode}}

\begin{figure}
 \begin{center}\includegraphics[width=5.5in]{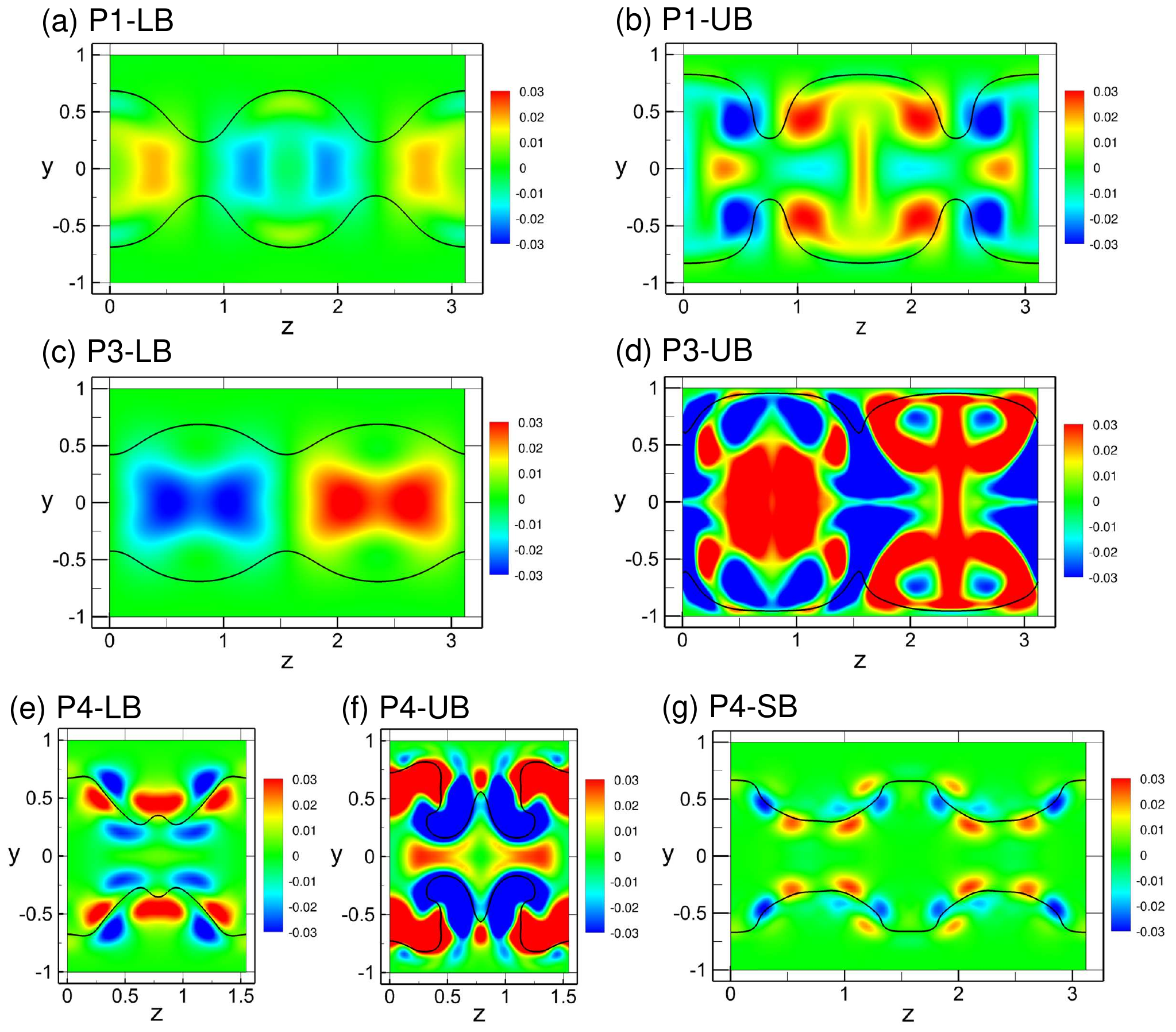}
  \caption{(Colour online) Contours of streamwise velocity deviations $u(x=x_c,y,z,t)-U(y,z)$ in the $y-z$ plane, where $U(y,z)$ is the streamwise-averaged streamwise velocity. {The streamwise location $x_{c}$ is chosen to illustrate the distinguishing features of the deviation.} The flow fields shown are (a) lower and (b) upper branch solutions for P1, (c) lower and (d) upper branch solutions for P3, (e) lower, (f) upper and (g) subharmonic branch solutions for P4. Except for the subharmonic solution, which is shown at $Re_c = 3600$, all solutions are at $Re_c = 1800$. The black line represents the critical layer in the $y-z$ plane at $x=x_c$. \label{fig:ufluc}}\end{center}
\end{figure}

{As described in the Introduction, prior work has addressed the structure and mechanism of nonlinear travelling waves in the context of nonlinear critical layer dynamics and in particular has noted the role that streamwise-wavy structures localized near the critical layer play in the self-sustaining process of at least one family of ECS \citep{Wang07,Hall10jfm}.
Therefore it is of interest to illustrate the channel flow ECS found here in relation to the critical layer position.} {In order to do so,} in figure \ref{fig:ufluc} we calculate velocity deviations $u(x=x_c,y,z,0)-U(y,z)$ in the $y-z$ plane of the P1, P3 and P4 lower and upper branch solutions along with the critical layer position (black thick line), where $u(x=x_c,y,z,0) = c_x$. Here $U(y,z)$ is the streamwise-averaged streamwise velocity (\emph{not} the streamwise- and spanwise-averaged velocity $U_{m}(y)$). {As the velocity deviation varies along the streamwise direction, a location $x_{c}$ in the streamwise direction is chosen so that the distinguishing features of the deviation are best illustrated.} {The full time-dependence of these structures is shown in supplementary movies 1-7.} The P1 and P3 lower branch solutions exhibit relatively strong deviations near the channel center as shown in figures \ref{fig:ufluc}(a) and (c), (and supplementary movies 1 and 3) from which observation we may call them ``core modes''. In particular, the P3 solution shows well-localized deviations near the channel center. Upper branch solutions exhibit stronger deviations throughout the channel height compared to lower branch solutions. In both cases, however, the fluctuations seem to be bounded between the top and bottom critical layers (also see accompanying online movies 2 and 4 in the supplementary materials {for figures \ref{fig:ufluc}(b) P1-UB and (d) P3-UB, respectively)}.

{P4 shows a different structure. Figures \ref{fig:ufluc}(e), (f) and (g) (and supplemental movies 5-7) show the streamwise velocity deviations for its lower and upper branch solutions at $Re_c = 1800$ and for a subharmonic solution at $Re_c = 3600$, respectively. The deviations of the lower and subharmonic branch solutions are highly localized very close to the critical layer, consistent with the Couette flow ECS results of \cite{Wang07} and \cite{Hall10jfm}. For the upper branch solutions, while strong deviations are observed across much of the channel, they are clearly organized by the critical layer. The clear organization of deviations around the critical layer suggests a connection to the critical layer dynamics, based on which this solution family may be called a ``critical layer mode''. {As seen in the vortical structures in figures \ref{fig:vortex}(h) and (i), the structure of P5 is also strongest around the critical layer.}}

This distinction between core modes and critical layer modes does not seem to have been previously identified. The lower branch Couette \citep{Blackburn13jfm,Hall10jfm,Wang07} and pipe \citep{Viswanath:2009da} flow travelling waves studied previously would be classified as ``critical layer'' rather than ``core'' modes, and the mechanistic studies of \cite{Blackburn13jfm} and \cite{Hall10jfm} are focused on critical layer modes.

In addition to movies for streamwise velocity deviations, there are additional accompanying online movies {8-17} in the supplementary material for the P1 - P5 travelling wave structures. These are shown in the $y-z$ plane, where streamwise velocity is represented by colour contours, wall-normal and spanwise velocities are shown by arrows and the critical layer is shown as a black curve.

\subsection{Exact coherent states on the laminar-turbulent boundary \label{sec:edge}}
Some of the lower branch travelling wave solutions in wall-bounded turbulent flows are embedded in the laminar-turbulent boundary \citep{Skufca06prl, Wang07}. If such solutions have only one unstable direction \citep{Skufca06prl}, which indicates that they are stable with respect to perturbations \emph{on} the boundary, they are edge states. By combining linear stability analysis of the nonlinear travelling wave families with direct time-integration of initial conditions perturbed along unstable directions of the travelling waves, we have determined whether the solutions live on the basic boundary and whether they are edge states. Specifically,  an initial condition for a trajectory is generated by addition of a small perturbation along an unstable eigenfunction to a lower branch solution. Both positive and negative perturbations are considered, and if there is an unstable direction for which the trajectory starting on one side of the lower branch solution develops into turbulence, while the other decays directly to the laminar state, then the solution lives on the basin boundary. If, additionally, there is only one unstable eigenvalue, then the travelling wave is attracting in all other directions besides its unstable one and is thus an edge state.

\begin{figure}
 \begin{center}\includegraphics[width=5.5in]{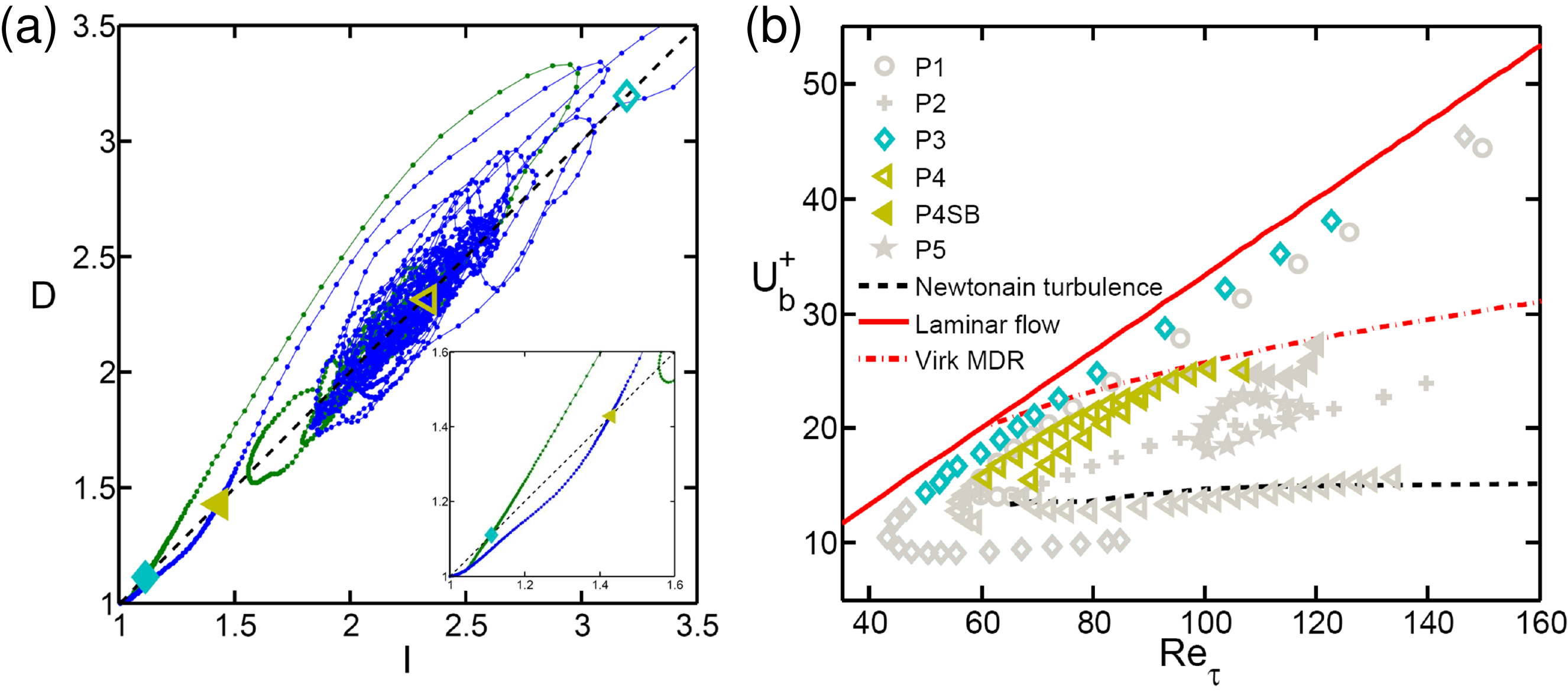}
  \caption{(Colour online) $(a)$ Time evolutions in the energy input rate and dissipation rate for DNS trajectories starting from nonlinear travelling waves  P3 ($\circ$) and P4 ($\diamond$) perturbed along an unstable eigendirection at $Re_{c} = 1800$. The solid and open symbols correspond to lower branch and upper branch states, respectively. Along trajectories, dot spacing is $\Delta t = 2$. The laminar state is at $(1,1)$. The dashed line represents $D=I$. (b) Bifurcation diagram with solutions on the basin boundary shown in colour. \label{fig:edge}}\end{center}
\end{figure}

Figure \ref{fig:edge}(a) shows dynamical trajectories along the unstable directions for {the P3 and P4 lower branch solutions at $Re_c=1800$}   projected onto the plane of energy dissipation rate ($\textit{D}$) and energy input rate ($\textit{I}$),
\begin{align}
& D = \frac{1}{2L_xL_z}\int_{0}^{L_z}\int_{-1}^{1}\int_{0}^{L_x}\left ( \left | \boldsymbol{\nabla} u \right |^2 + \left | \boldsymbol{\nabla} v \right |^2 +\left | \boldsymbol{\nabla} w \right |^2 \right )\textrm{d}x\textrm{d}y\textrm{d}z, \\
& I = \frac{1}{2L_z}\int_{0}^{L_z}\int_{-1}^{1}\left ( pu|_{x=0} -pu|_{x=L_x}\right )\textrm{d}y\textrm{d}z.
\end{align}
Both values are normalized by their laminar values such that the laminar state is at $(1, 1)$. Recall that the fundamental domain size is different between P3 and P4: $L_x = 2\pi$ and $L_z = \pi$ for P3 and $L_x = \pi$ and $L_z = \pi/2$ for P4. In both cases, one perturbation (the one that increases $I$ and $D$) leads to turbulence and the other to laminar, indicating that these travelling waves are on the basin boundary for turbulent flow in their respective domains. {Even though there are multiple unstable eigenvalues for the P4 lower and subharmonic branches, only one (real) unstable eigenvalue gives the aforementioned two types of trajectories of perturbations.} Figure \ref{fig:edge}(b) shows the results of this analysis for the entire bifurcation diagram -- the solutions shown in gray are \emph{not} on the basin boundary while other others are -- we find that only P3 and P4 display parameter ranges where they are on the boundary. For P3 this range is  $50 \lesssim Re_{\tau} \lesssim  123~(1200 \lesssim Re_c \lesssim 5000)$. For P4, both lower and subharmonic branch solutions are found to lie on the basin boundary for their respective domains. {The ranges are $69 \lesssim Re_{\tau} \lesssim 88.7~(1600 \lesssim Re_c \lesssim 3070)$ for P4-LB, $61 \lesssim Re_{\tau} \lesssim 88.7~(1400 \lesssim Re_c \lesssim 3070)$ for P4-LB2, and $88.9 \lesssim Re_{\tau} \lesssim 107~(3080 \lesssim Re_c \lesssim 4000)$ for P4-UB}. Furthermore, since it is found that the P3 lower branch solutions have only one unstable eigenvalue {in the symmetric subspace}, they are indeed edge states {in that space}. Even though there is only one eigenfunction that gives rise to an escape scenario from the basin boundary for the P4 lower branch solutions, they have multiple unstable eigenvalues so are not edge states.

\subsection{Connections between travelling wave solutions and turbulent trajectories \label{sec:connect}}
One motivation for studying nonlinear travelling waves in shear flows is the idea that these states form the state-space skeleton of the turbulent dynamics. In this section, we address this issue, examining how close the turbulent trajectories approach the travelling wave solutions. We focus on the domain $L_x \times L_y \times L_z = \pi \times 2 \times \pi/2$, the same box size as the P4 solution family. We choose {$Re_c = 1800~ (Re_{\tau} = 85)$} and perform simulations at constant mass flux. {In contrast to the travelling wave computations, these turbulence simulations are performed without imposing any symmetries on the flow.}  Comparisons are made with travelling waves that have the same $Re_c$. Two methods of comparison are used: the first examines the probability distribution function for the instantaneous mean velocity profile and the second projects the state space dynamics into three dimensions corresponding to physically meaningful averaged quantities.

\begin{figure}
 \begin{center}\includegraphics[width=5.5in]{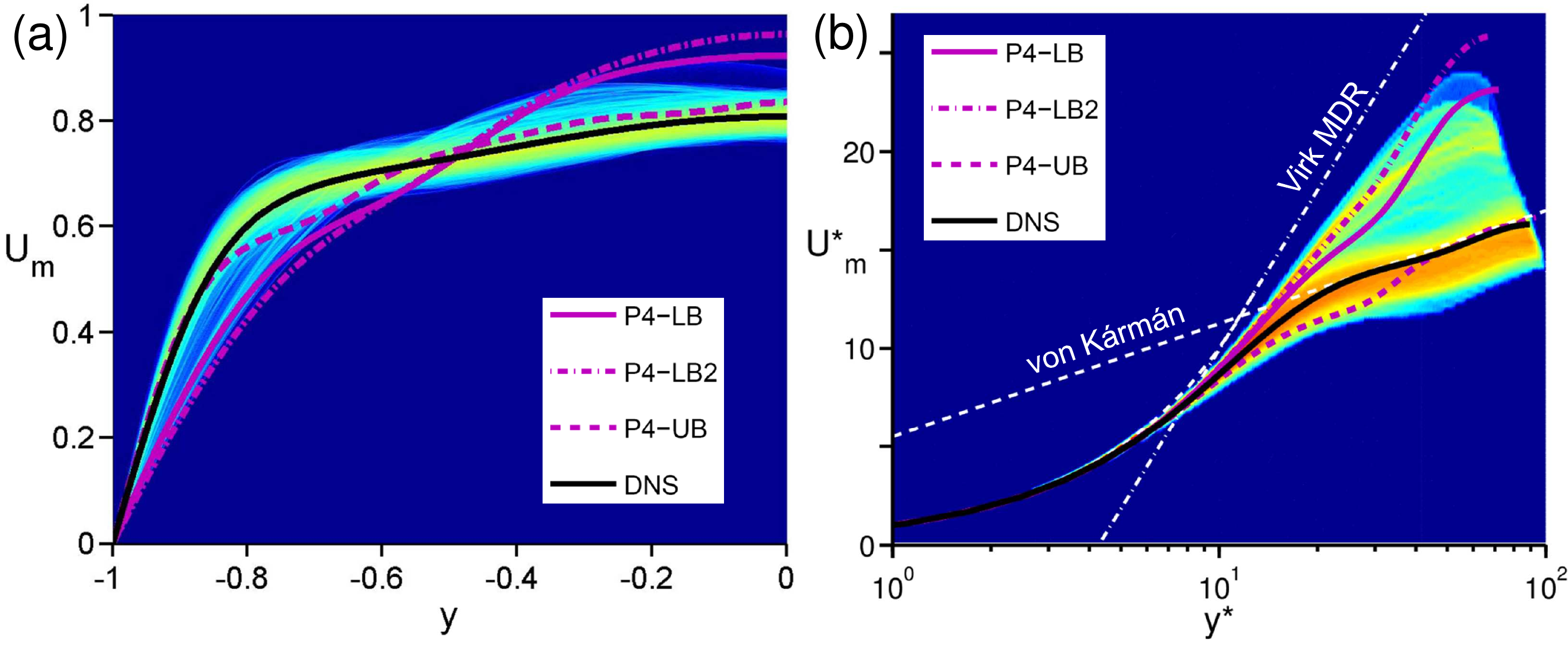}
  \caption{(Colour online) The probability density function of mean velocity profile from DNS, along with P4 lower branch (P4-LB), another lower branch (P4-LB2), and upper branch (P4-UB) solutions on (a) {outer scaling} and (b) '$\ast$' scaling (inner units based on instantaneous area-averaged wall shear stress). The black line is the time-averaged DNS velocity profile. A logarithmic scale is used, with blue indicating vanishing probability. \label{fig:umpdf}}\end{center}
\end{figure}

Figures \ref{fig:umpdf}(a) and (b) show the probability density functions (PDFs), plotted on a logarithmic scale, for the mean velocity profiles {at each wall normal position $y$ (figure \ref{fig:umpdf}(a)) or $y^{*}$ (figure \ref{fig:umpdf}(b))} in DNS based on outer units and `$\ast$'-scaling (instantaneous inner scales), respectively. {The PDF is normalized so that the integral over the whole PDF equals 1.} {Using outer scaling (non-time-dependent scaling), it is difficult to compare a DNS trajectory to TW solutions because each TW has different friction velocity. However,} as highlighted by previous studies \citep{Xi:2012io, Agostini14pof}, the `$\ast$'-scaling, which leads all profiles to collapse to the same curve near the wall, is the proper one to use for instantaneous quantities with which a TW solution can be directly compared to an instantaneous flow field. {Here, we used DNS results for 40000 time units to compute PDFs. According to \cite{Xi:2012io} and our calculations, approximately 8 -- 9\% of the total simulation time is spent near the Virk-like state. Thus the data is sufficient to capture approaches to the Virk log-law in PDFs.} In figure \ref{fig:umpdf}(a) we see that near the wall, the DNS velocity profile is very nearly bracketed between the P4 upper branch and lower branch solutions, while deviations from these solutions become more prevalent near the center. The same trend is apparent in the plot in instantaneous inner units, figure \ref{fig:umpdf}(b), which emphasizes the strong similarities in the near-wall behaviour as well as the transient approaches of the DNS mean velocity profile toward the P4 lower branch solutions. It appears that for $y^{*}\lesssim 30$ the P4 travelling waves form an approximate envelope for the PDF of the DNS mean velocity profile. Furthermore, relatively high probability regions (red) are observed around the von \Karman{} log-law and P4-UB solution. Interestingly, there is also a slightly high probability region (yellow) close to the P4-LB solutions.

\begin{figure}
 \begin{center}\includegraphics[width=5.0in]{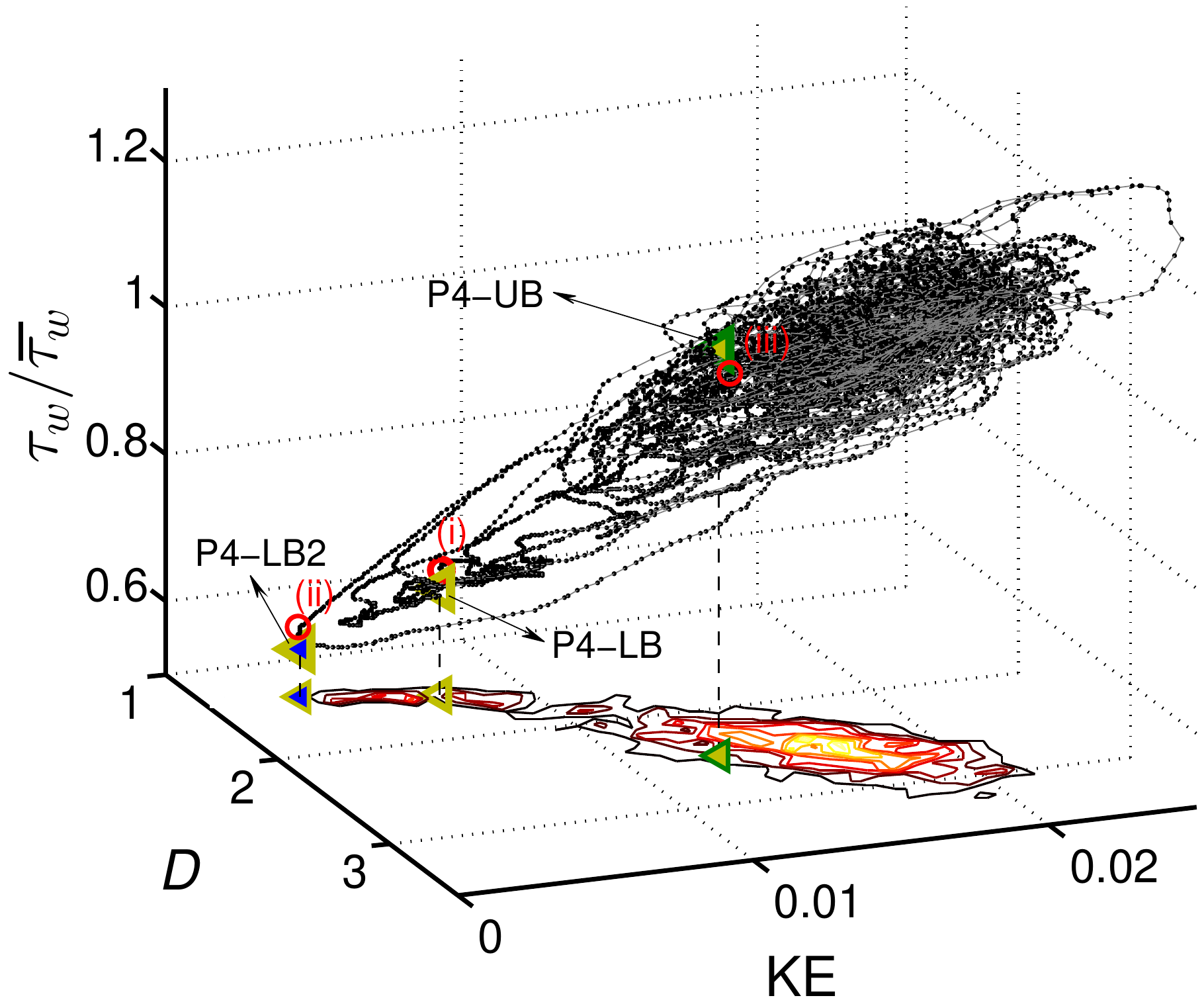}
  \caption{(Colour online) A state-space visualization of DNS trajectories, projected onto three dimensions: disturbance kinetic energy (KE), an energy dissipation rate (\textit{D}), and normalized instantaneous wall shear stress ($\tau_w/\bar{\tau}_w$). The grey line indicates the turbulent trajectory, to which black dots are attached at intervals of $1 h/U_c$. A joint probability function of KE and \textit{D} is shown at the bottom of the figure. The labelled symbols ($\triangleleft$) are P4 solutions. Points (i), (ii), and (iii) are the closest visits to P4-LB, P4-LB2, and P4-UB, respectively. Note that all quantities are calculated only for the  bottom half of the channel. \label{fig:state}}\end{center}
\end{figure}

\begin{figure}
 \begin{center}\includegraphics[width=5.5in]{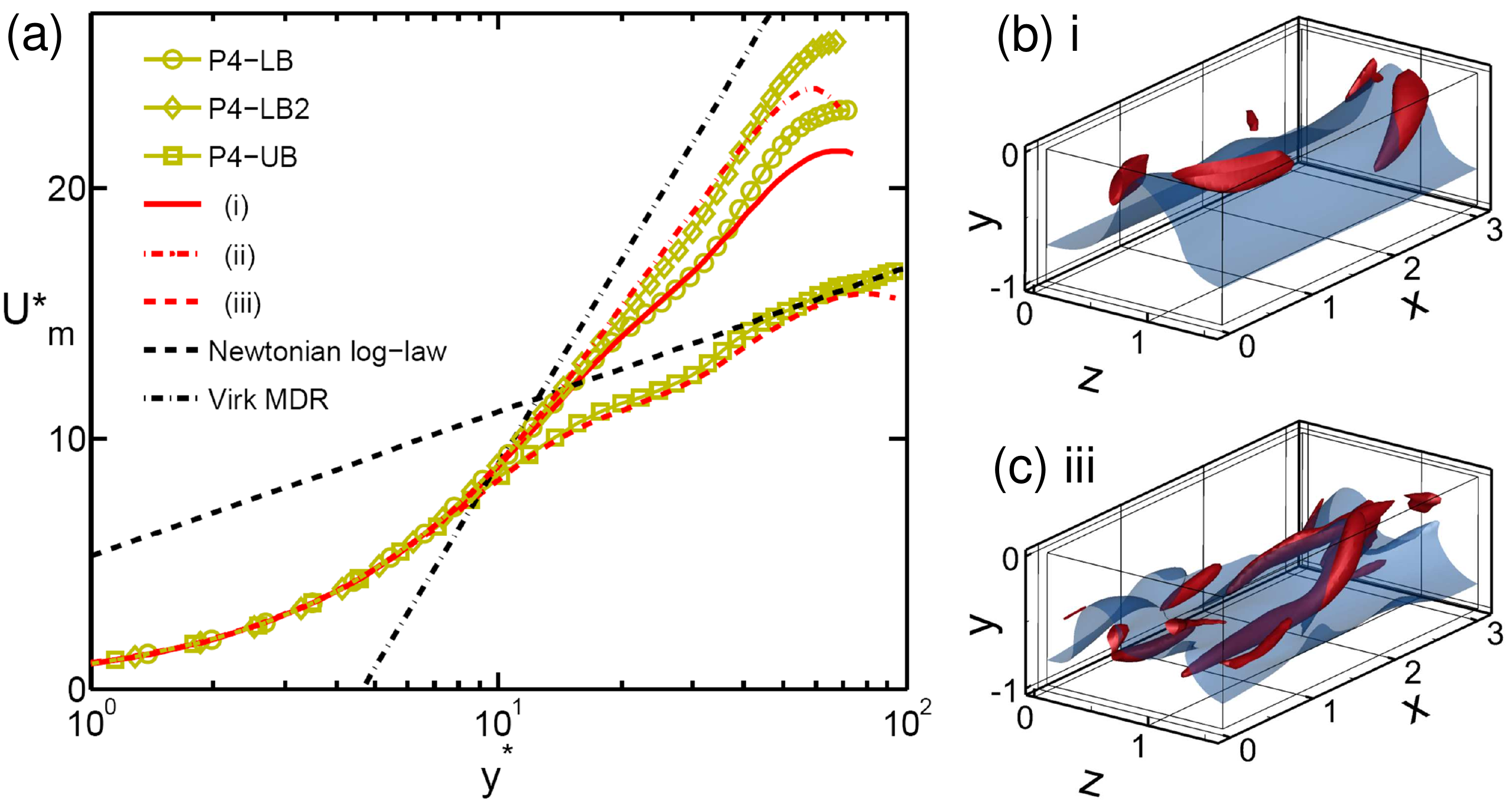}
  \caption{(Colour online) (a) Instantaneous mean velocity profiles for {instants (i), (ii), and (iii), where the closest approach to P4 lower branch (LB), second lower branch (LB2), and upper branch (UB) solutions are observed, respectively. (b)-(c) Flow structures at instants (i) and (iii)}.
\label{fig:compare}}\end{center}
\end{figure}

Now we visualize the approach of turbulent trajectories to travelling wave solutions in state space. To do so, we project turbulent trajectories onto a three-dimensional space using the following quantities: disturbance kinetic energy (KE), energy dissipation rate (\textit{D}), and area-averaged instantaneous wall shear stress normalized by its mean value ($\tau_w/\bar{\tau}_w$). The disturbance kinetic energy is defined as follows:
\begin{equation}
\textrm{KE} = \frac{1}{L_xL_z}\int_0^{L_z}\int_{-1}^{0}\int_0^{L_x}\frac{1}{2}\left(\boldsymbol{u}-\boldsymbol{u}_{lam} \right)^2 ~\textrm{d}x~\textrm{d}y~\textrm{d}z
\end{equation}
where $\boldsymbol{u}_{lam}$ is the parabolic laminar profile. Figure \ref{fig:state} shows a turbulent trajectory as well as the P4 travelling waves projected onto these three quantities, as well as the joint probability density function (PDF) of KE and \textit{D} at the bottom of the figure. Note that all quantities are calculated for only the bottom half of the domain. The dynamical trajectory spends most of its time within one core region of state space, which we can identify with normal or ``active'' turbulence; the  P4-UB solution is in this region, as also seen on the joint PDF. The trajectory occasionally escapes, however, from the active region, approaching the P4 lower branch solutions. During these excursions, some trajectories pass through the vicinity of P4-LB, approaching P4-LB2 very closely. When returning to the active region, the path strongly overshoots the core of these region and might be considered as a turbulent burst. {Similar observations on the relationship between invariant states and bursting events have been made for Couette flow \citep{Kawahara01jfm} and channel flow \citep{Toh03jfm}.} The closest visits to P4-LB, P4-LB2, and P4-UB, respectively, are labelled as points (i), (ii), and (iii). The mean velocity profiles at these three instants and for the P4 travelling waves are plotted in figure \ref{fig:compare}(a). {The profiles for instants (i) and (ii) appear very similar to the P4-LB and P4-LB2 solutions, respectively, while instant (iii) has a similar profile to the P4 upper branch and the von \Karman{} mean profile.} Flow structures for  instants (i) and (iii) are visualized in figures \ref{fig:compare}(b) and (c), where to facilitate comparison we use the same vortex strength and critical layer isosurfaces as in figures \ref{fig:vortex}(f) P4-LB and (g) P4-UB. Note that flow structure for instant (ii) is very similar with that of instant (i), but shows weaker vortex motion. There is substantial similarity between the vortical and critical layer structures of the snapshots and the travelling waves. The critical layer for instant (i) shows weak streamwise variation and the vortex motions are seen to be localized around the critical layer as also seen for P4-LB. Similarly, instant (iii) and P4-UB resemble one another, particularly with regard to their inclined vortical structures.

\begin{figure}
 \begin{center}\includegraphics[width=4.7in]{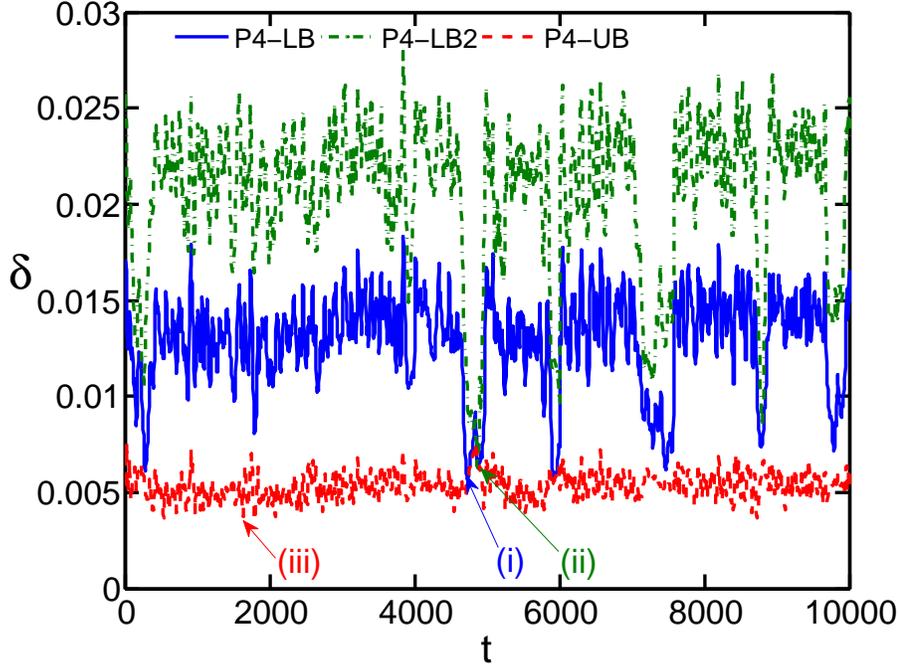}
  \caption{(Colour online) Distances $\delta$ between a DNS trajectory and P4 travelling wave solutions. Instants (i), (ii), and (iii) are the closest visits to P4-LB, P4-LB2, and P4-UB, respectively, indicated on figure~\ref{fig:state}. \label{fig:distance}}\end{center}
\end{figure}

{To further address the relationship between the DNS trajectories and travelling waves, we calculated a norm of the difference between a DNS velocity field $\boldsymbol{u}$ and a travelling wave.
Taking into account the arbitrary phase in $x$ and $z$ of the velocity fields, this distance, which we denote $\delta$, is calculated as follows:
\begin{equation}
\delta(t) = \min_{0\leqslant x_0 < L_x}~\min_{0\leqslant z_0 < L_z} \parallel \boldsymbol{u}(x+x_0,y,z+z_0,t)-\boldsymbol{u}_{\textrm{TW}}(x,y,z) \parallel,
\end{equation}
where $x_0$ and $z_0$ are phase shifts in the $x$ and $z$ directions, respectively. Time series of $\delta(t)$ computed for the  P4 solution family are shown in figure \ref{fig:distance}. The distances between P4-LB and instant (i), P4-LB2 and instant (ii), and P4-UB and instant (iii) are $5.7\times 10^{-3}$, $6.2\times 10^{-3}$, and $3.5\times 10^{-3}$, respectively. Those values are comparable to distances at the points of the closest visits to travelling waves for pipe flow \citep{Viswanath09jfm} and to equilibria for Couette flow \citep{Halcrow:2009jx}, which are of order $O(10^{-3})$. Thus, the closeness of turbulent trajectories to P4 travelling waves is identified using full velocity fields for minimal channel flow.}

{The state-space picture that clearly shows close approaches to multiple TW solutions and the closeness to the TW solutions using full velocity fields have yet to be reported in the channel flow literature. Furthermore, it must be emphasized that those multiple travelling waves belong to the same solution family.}

These results confirm the hypothesis posed in prior work \citep{Graham:2014uj,Xi:2010em,Xi:2010fv,Xi:2012io} that the ``active'' and ``hibernating'' phases of minimal channel turbulence correspond to time intervals where the trajectory is close to upper and lower branch travelling waves, respectively. Finally, returning to the bifurcation diagram we recall that there are also two upper branch P4 solutions, but we were only successful in computing one of them over a broad range of $Re_{c}$. We might speculate that if the second upper branch solution could be found at the $Re_{c}$ at which we have performed the DNS it would also lie in the core active turbulence region.

\section{Conclusion \label{sec:conclusion}}
We have computed five new families of nonlinear travelling wave solutions, denoted P1-P5, in Newtonian plane Poiseuille flow.  As Reynolds number is increased, the P1 and P3 lower branch solutions become close and parallel to the laminar solution branch, indicating that they are very low-drag states. The P2 solution branch results from a symmetry-breaking bifurcation from P1. P5 forms a closed loop (isola).  Most interestingly, the P4 solution family shows very intriguing behaviour in terms of mean properties as Reynolds number is increased. The average velocities of the lower and upper branches appear to approach the Virk MDR profile observed in viscoelastic turbulence and the classical Newtonian (von \Karman{}) profiles, respectively. {The former observation adds to the set of results in which mean velocity profiles close to the Virk profile are found in Newtonian flow \citep{Kerswell2003pof,Bandyopadhyay:2006ji,Xi:2010em,Xi:2010fv,Dubief:2011uo, Xi12a,Xi:2012io}.} On the lower branch, a subharmonic bifurcation arises around $Re_{\tau} \approx 90$, giving rise to spatiotemporal period doubling. 

The structures and symmetries of the various solution families are described. The fluctuations of the P1 and P3 solutions are largest near the channel center so we have denoted them as core modes, while the P4 and P5 solutions display fluctuations localized around the critical layer so we call them critical layer modes. Over a range of Reynolds numbers P3 and P4  lower branch solutions are embedded in the laminar-turbulent boundary.

Finally, we addressed the issue of how close the turbulent trajectories approach the travelling wave solutions, focusing on the P4 family. In prior work \citep{Graham:2014uj,Xi:2010em,Xi:2010fv,Xi:2012io}  it was hypothesized that ``active'' and ``hibernating'' phases of minimal channel turbulence correspond to time intervals where the trajectory is close to upper and lower branch travelling waves, respectively. The present results corroborate this hypothesis. The turbulent trajectory spends most of its time within a region of state space that can be identified as normal or ``active'' turbulence; the P4-UB solution is in this region, while the hibernating intervals are approaches to the P4 lower branch solutions.  


\section*{Acknowledgements}
The authors thank Kengo Deguchi, John Gibson, Rich Kerswell and Masato Nagata for illuminating discussions and especially Kengo Deguchi and John Gibson for providing their travelling wave solutions. This work has been supported by the Air Force Office of Scientific Research through grants FA9550-11-1-0094 and FA9550-15-1-0062 (Flow Interactions and Control Program) and by the National Science Foundation through grants CBET-1066223 and CBET-1510291 (Fluid Dynamics program). The direct numerical simulation code used here was developed and distributed by John Gibson at the University of New Hampshire.

\bibliographystyle{jfm}
\bibliography{refs,turbulence-MDG}
\end{document}